# Gravitational Radiation from First-Order Phase Transitions

Marc Kamionkowski*

*School of Natural Sciences, Institute for Advanced Study, Princeton, NJ 08540*

Arthur Kosowsky[†] and Michael S. Turner[‡]

*NASA/Fermilab Astrophysics Center, Fermi National Accelerator Laboratory, Batavia, IL 60510-0500*
*and*
*Departments of Physics and of Astronomy & Astrophysics*
*Enrico Fermi Institute, The University of Chicago, Chicago, IL 60637-1433*

(October, 1993)

## Abstract

We consider the stochastic background of gravity waves produced by first-order cosmological phase transitions from two types of sources: colliding bubbles and hydrodynamic turbulence. First we discuss the fluid mechanics of relativistic spherical combustion. We then numerically collide many bubbles expanding at a velocity $v$ and calculate the resulting spectrum of gravitational radiation in the linearized gravity approximation. Our results are expressed as simple functions of the mean bubble separation, the bubble expansion velocity, the latent heat, and the efficiency of converting latent heat to kinetic energy of the bubble walls. A first-order phase transition is also likely to excite a Kolmogoroff spectrum of turbulence. We estimate the gravity waves produced by such a spectrum of turbulence and find that the characteristic amplitude of gravity waves produced is comparable to that from bubble collisions. Finally, we apply these results to the electroweak transition. Using the one-loop effective potential for the minimal electroweak model, the characteristic amplitude of gravity waves produced is $h \simeq 1.5 \times 10^{-27}$ at a characteristic frequency of $4.1 \times 10^{-3}$ Hz corresponding to $\Omega \sim 10^{-22}$ in gravity waves, far too small for detection. Gravity waves from more strongly first-order phase transitions, including the electroweak transition in non-minimal models, have better prospects for detection, though probably not by LIGO.

---

*kamion@guinness.ias.edu.

[†]NASA GSRP Fellow. arthur@oddjob.uchicago.edu

[‡]mturner@fnalv.fnal.gov

04.30.+x, 98.70.Vc, 98.80.Cq

# I. INTRODUCTION

First-order phase transitions in the early Universe can be potent sources of gravitational radiation [1,2]. In a recent series of papers we have calculated the radiation emitted by colliding vacuum bubbles and obtained useful approximations to the bubble dynamics, and applied these results to very strongly first-order phase transitions which occur through nucleation and percolation of vacuum bubbles [3–5]. In this paper, we extend these results to more weakly first-order phase transitions which occur in a thermal environment, and apply our results to the electroweak phase transition.

In a first-order phase transition, the Universe starts in a metastable high-temperature phase (the "symmetric" phase) and converts to a stable low-temperature (the "broken-symmetry") phase. The transition proceeds via nucleation of bubbles of the low-temperature phase within the high-temperature phase; these bubbles then expand and merge, leaving the Universe in the broken-symmetry phase.

Previously, we considered vacuum transitions, in which the only component of the Universe is a scalar field. In this case true-vacuum bubbles are nucleated through quantum tunneling [6]. The dynamics of these bubbles is comparatively simple: once the bubbles are nucleated, the scalar field simply evolves according to the Klein-Gordon equation. Bubbles that are larger than a critical size begin to expand and rapidly approach velocities near the speed of light. All of the liberated vacuum energy goes into accelerating the bubble walls, which become progressively thinner and more energetic as the bubbles expand. These high velocities and large energy densities provide the necessary conditions for generating large amounts of gravitational radiation, and the resulting radiation spectrum depends very simply on the natural length and energy scales of the problem.

For a thermal transition, the problem is more complex. Nucleation of bubbles of the low-temperature phase occurs through quantum tunneling and thermal fluctuations. However, the evolution of these bubbles is not driven simply by scalar-field evolution. Instead, it depends on interactions of the bubble wall with the plasma and on the resulting fluid dynamics. Part of the latent heat released in the transition raises the plasma temperature, while another fraction of the latent heat is converted to bulk motions of the fluid. If the Reynolds number of the universe at the phase transition is large enough, then bubble motion produces turbulence in the plasma.

In this paper, we perform detailed calculations of the gravitational radiation produced by the collision of spherical combustion bubbles expanding at a velocity $v$, using the linearized gravity approximation. The resulting spectra are simply expressed in terms of $v$, the logarithmic derivative of the bubble-nucleation rate $\beta$, the ratio of vacuum to thermal energy density $\alpha$, and an efficiency factor $\kappa$ giving the fraction of vacuum energy which goes into kinetic energy of bulk motions of the fluid, as opposed to heating. As discussed below, the theory of relativistic combustion gives $\kappa$ and $v$ as a function of $\alpha$, which in effect, measures the degree of supercooling (*i.e.*, how strongly first order the phase transition is).

Combustion occurs via two distinct modes: detonation and deflagration. Roughly, detonations occur when the phase boundary propagates faster than the speed of sound, while for deflagrations the phase boundary propagates slower than the sound speed. We show that the bubble collisions in phase transitions proceeding via detonation will produce substantial gravitational radiation. In contrast, production of gravitational radiation from collisions of deflagration bubbles should be small, because the bubble velocities are small (subsonic). It has recently been argued that detonation is the only stable mode of combustion for a cosmological phase transition, and that a transition which begins via deflagration rapidly becomes unstable and converts to detonation [7]. For these reasons we mainly focus on gravity waves produced by detonation bubble collisions.

Both modes of combustion can stir up turbulence on scales comparable to the bubble size. We estimate the gravity waves produced by a fully developed Kolmogoroff spectrum of turbulence through simple dimensional arguments, and find that the amplitude of the spectrum is comparable to that from bubble collisions. This source will generate gravity waves in addition to those produced by the actual bubble collisions. We note that our estimates are completely general, and apply to any injection of energy in the early Universe on a large length scale.

Section II discusses the relevant results from relativistic combustion theory. We review the solutions to the hydrodynamic equations of motion for spherically symmetric detonation bubbles [8] and derive relationships between bubble-expansion velocity, bubble kinetic energy, latent heat, and temperature. We also discuss the solutions for spherical, relativistic deflagration bubbles, which have not been previously addressed, and briefly compare with the hydrodynamics associated with planar combustion [9–11]. In Section III, we review gravity-wave formalism used for our calculations. The calculation of the gravitational radiation produced by many colliding bubbles is made tractable through the envelope approximation [5]; we discuss the applicability of this approximation to combustion bubbles. Then we numerically calculate the radiation spectra for the collision of many bubbles in terms of their expansion velocity and kinetic energy, which are related to parameters of the phase transition in Section II. Estimates of gravity waves from turbulence conclude the section. Section IV contains the necessary formulas to propagate the generated spectrum of gravity waves to the present time. As a sample application, we derive the gravitational radiation produced by the electroweak transition, using the one-loop effective potential of the minimal standard model. We conclude by briefly considering detection prospects, especially for more strongly first-order phase transitions. Appendix A analyzes spherical relativistic deflagration bubbles, and in Appendix B a model effective potential is analyzed and applied to the electroweak transition.

## II. FLUID FLOW IN SPHERICAL COMBUSTION

In order to calculate the spectrum of gravitational radiation from colliding bubbles, we need to know the spatial components of the traceless part of the stress-energy tensor, $T_{ij}$. For a relativistic fluid, this is simply $T_{ij} = w\gamma^2 v_i v_j$, where $w = e + p$ is the enthalpy density, $e$ and $p$ are the energy density and pressure, $v_i$ are the components of the fluid velocity, and $\gamma = (1 - |\mathbf{v}|^2)^{-1/2}$ is the Lorentz factor. For spherical bubbles, the only nonvanishing component of the stress tensor is $T(r) \equiv T_{rr}$, and the fluid velocity has only a radial component $v \equiv v_r$. The radial dependence of the enthalpy density $w(r)$ and fluid velocity $v(r)$ need to be determined. Gravitational radiation from thin-wall bubbles depends on the quantity

$$\int T(r)r^2 dr = \int w(r)\frac{v(r)^2}{1-v(r)^2}r^2 dr. \tag{1}$$

The rest of this Section is devoted to evaluating this expression.

To model a phase transition, we assume that the equation of state of the gas in the high-temperature ("symmetric" or "unburnt") phase describes a relativistic gas plus a false-vacuum energy density:

$$e_1 = aT_1^4 + \epsilon, \tag{2}$$
$$p_1 = \frac{1}{3}aT_1^4 - \epsilon, \tag{3}$$

where $\epsilon$ is the false-vacuum energy density (or equivalently, 1/4 of the latent heat). In the low-temperature ("broken" or "burnt") phase the equation of state is simply that for a relativistic gas:

$$e_2 = aT_2^4, \tag{4}$$
$$p_2 = \frac{1}{3}aT_2^4. \tag{5}$$

Note that $w_i = (4/3)aT_i^4$. When a bubble forms in a first-order transition, its interior is described by the broken phase equation of state, while its exterior is in the symmetric phase. The phase boundary at the bubble wall, the "detonation front", is assumed to be infinitely thin. The difference in free energy between the inside and the outside of the bubble creates an effective pressure driving the expansion of the bubble. We define the quantity

$$\alpha = \epsilon/aT_1^4, \tag{6}$$

the ratio of vacuum energy to the thermal energy in the symmetric phase; $\alpha$ characterizes the strength of the phase transition. The limits $\alpha \to 0$ and $\alpha \to \infty$ correspond to very weak and very strong first-order phase transitions, respectively.

In spherical combustion there is no natural length scale, and the hydrodynamic equations can be written in terms of $\xi = r/t$ where $r$ is the distance from the center of the bubble and $t$ is the time since nucleation. In other words, the velocity and enthalpy-density profiles, $v(r,t)$ and $w(r,t)$, are self-similar, being functions of only $r/t$. The variable $\xi$ is then the outward velocity of a given point in the bubble profile. As shown by Steinhardt [8], Euler's equation and the equations of continuity and conservation of entropy can be combined in the case of spherically symmetric flows to yield an equation for the radial velocity as a function of $\xi$:

$$\gamma^2(1-v\xi)\left[\left(\frac{\mu}{c_s}\right)^2 - 1\right]\frac{dv}{d\xi} = \frac{2v}{\xi}, \tag{7}$$

where $\mu \equiv (\xi - v)/(1 - v\xi)$ and $\gamma^2 = (1-v^2)^{-1}$. The enthalpy density satisfies

$$\frac{1}{w}\frac{dw}{dv} = \frac{4\gamma^2\mu}{3c_s^2}, \tag{8}$$

which can be integrated in terms of the velocity profile:

$$w(\xi) = w_d \exp\left[-\frac{4}{3c_s^2}\int_{v(\xi)}^{v_d} \gamma^2 \mu dv\right]. \tag{9}$$

The stress-tensor $T(r)$ can then be obtained from the solutions to these equations with the proper boundary conditions.

Conservation of energy and momentum assure that *in the rest frame of the bubble wall*, $\beta_1$, the velocity of fluid in the symmetric phase into the wall, is given by [8,9,12]

$$\beta_1 = \left[\frac{(p_2-p_1)(e_2+p_1)}{(e_2-e_1)(e_1+p_2)}\right]^{1/2}, \tag{10}$$

and that $\beta_2$, the velocity of fluid in the broken-symmetry phase away from the wall, is

$$\beta_2 = \left[\frac{(p_2-p_1)(e_1+p_2)}{(e_2-e_1)(e_2+p_1)}\right]^{1/2}. \tag{11}$$

The enthalpy densities on each side of the wall are related by

$$\frac{w_1\beta_1}{1-\beta_1^2} = \frac{w_2\beta_2}{1-\beta_2^2}. \tag{12}$$

If $w_1$ (*i.e.*, the temperature outside) and $\alpha$ are given, $\beta_1$, $\beta_2$ and $w_2$ are still undetermined; however, once one of the three quantities is given, the other two are determined.

It has been shown [9] that there are two qualitatively different kinds of combustion. If $\beta_1 > \beta_2$, the transition occurs via "detonation" and the wall propagates at a supersonic velocity, *i.e.*, at a velocity larger than $c_s$, the speed of sound; if $\beta_1 < \beta_2$, the transition occurs via "deflagration," and the wall propagates at subsonic velocity. The sound velocity is given by $dp/de$ at constant entropy; in general, it is a function of the state variables, but in the highly relativistic limit $c_s \to 1/\sqrt{3}$. In the remainder of this paper we always take this limiting value for the sound velocity. It has also been shown [8] (and will be discussed below) that if the transition occurs via detonation, $\beta_2 = c_s$ and so $\beta_1$ and $w_2$ are given simply in terms of $\alpha$ and $w_1$. However, for deflagrations, $\beta_2$ is, in general, still undetermined.

In either case, the fluid velocities (in the rest frame of the wall) in and out of the wall are unequal, $\beta_1 \neq \beta_2$, so the fluid velocity $v$ must be nonzero somewhere. Moreover, the fluid velocity is zero at the center of the bubble (by spherical symmetry) and far away from the bubble (in the "rest" frame of the Universe). Thus, deflagration or detonation is characterized by a radial fluid velocity profile, $v(r)$, which satisfies the fluid Eqs. (7) and (8) with the appropriate boundary conditions. We now discuss the solution to this hydrodynamic problem.

### A. Detonations

The case of detonations has been discussed in detail by Steinhardt [8], and we review the relevant results here. If the transition proceeds via detonation, the unburnt fluid enters the wall at a supersonic velocity. For this reason, there can be no shock preceding the wall, so the fluid is at rest outside the bubble wall; i.e., $v(\xi) = 0$ for $\xi > \xi_d$ where $\xi_d = \beta_1$ is the propagation velocity of the wall. Since $\beta_1 > \beta_2$, the fluid just behind the detonation front is accelerated outward to a velocity $(\xi_d - \beta_2)/(1 - \xi_d \beta_2)$ (this is just the relativistic transformation of the velocity from the wall frame to the rest frame of the bubble). As shown by Steinhardt, the detonation front is then followed by a rarefaction wave in which the velocity profile $v(\xi)$ falls smoothly to zero at $\xi = c_s$, and remains zero for $\xi \leq c_s$.

Steinhardt also showed that detonation solutions to Eq. (7) exist only if $\beta_2 = c_s$. This is the relativistic generalization of the Chapman-Jouget condition for spherical detonations (see Ref. [12]). For a general planar detonation [9–11], the value of $\beta_2$ is not constrained to be $c_s$. Therefore, the detonations in a phase transition in the early Universe, restricted to satisfy the Chapman-Jouget condition, are not as general as those considered in some previous work [9–11]. We should also point out that the functional form of the velocity and enthalpy-density profiles are different in a spherical detonation from those in a planar detonation (even with the Chapman-Jouget condition), although they are similar qualitatively.

Given $\beta_2 = c_s$, one finds that the velocity of the detonation front, $\xi_d$, for a given $\alpha$ is simply [8]

$$\xi_d = \frac{1/\sqrt{3} + (\alpha^2 + 2\alpha/3)^{1/2}}{1 + \alpha}. \tag{13}$$

In Fig. 1 we plot the velocity of propagation of the detonation front, $\xi_d$, as a function of $\alpha$, the parameter describing the strength of the transition. The velocity profile is then given by integrating Eq. (7) with the boundary condition $v(\xi_d) \equiv v_d = (\xi_d - c_s)/(1 - \xi_d c_s)$, from $\xi = \xi_d$ to $\xi = c_s$.

As $\xi \to \xi_d$, $dv/d\xi \to \infty$, so Eq. (7) cannot be easily integrated numerically from $\xi = \xi_d$. Instead, we write $\xi$ as a function of $v$, use the relation $\xi(v) \simeq \xi_d + (1/2)\xi''(v_d)$ for $v \to v_d$, and integrate from some $v$ very close to $v_d$. Here,

$$\xi''(v_d) = \frac{-\xi_d \gamma_d^2 (\xi_d c_s - 1)^2}{(\xi_d - c_s) c_s} \tag{14}$$

is the second derivative of $\xi$ with respect to $v$ at the detonation front. The velocity profiles for several values of $\alpha$ are displayed in Fig. 2. As shown, $v(\xi)$ is zero for $\xi < c_s$; there is a weak discontinuity at $\xi = c_s$, and $v$ increases until $\xi = \xi_d$ where $dv/d\xi \to \infty$. Also, as $\alpha$ is increased, both $\xi_d$ and $v_d$ increase.

Once the velocity profile has been determined, the enthalpy-density profile can be calculated using Eq. (9). The enthalpy density at the detonation front, $w_d = w_2$, can be determined in terms of $w_1$ and $\alpha$ from Eq. (12). Numerically integrating Eq. (9) is straightforward, but as the detonation becomes strong, $(\alpha \gtrsim 1)$, $w(\xi)$ varies rapidly near the detonation front. The quantity $\mu(\xi)$ equals $c_s$ at $\xi = c_s$, increases until some $\xi$ which becomes closer to $\xi_d$ as $\alpha$ is increased, and then rapidly decreases to $c_s$ at the front. One finds that the region near $\xi_d$ where $\mu$ is decreasing loosely defines a width—which decreases as $\alpha$ is increased—for the detonation front, and that $w(\xi)$ varies quite rapidly in this region. The enthalpy-density profile, $w(\xi)$, divided by $w_1$, the enthalpy density outside the bubble, is plotted in Fig 3. The enthalpy density jumps at the detonation front, then decreases smoothly until $\xi = c_s$ and maintains a constant value $w_0 < w_1$ at the center of the bubble, $\xi < c_s$. For larger $\alpha$, $w(\xi)$ becomes increasingly concentrated near the detonation front.

In Fig. 4, we plot the stress-energy density $T(\xi) = wv^2\gamma^2$. Note that as $\alpha \to 0$, all the stress-energy becomes concentrated near a thin shell near $\xi = c_s$, while as $\alpha$ is increased, the stress-energy becomes dramatically concentrated near the detonation front. The thickness of this shell tends to zero in both the strong- and weak-detonation limits and always remains negligible compared with the bubble radius; thus a thin-wall approximation to the stress-energy tensor of a detonation bubble is valid.

To gauge the accuracy of our numerical integration, we checked that the energy contained in a volume of equivalent size before the bubble was nucleated, $E_{\text{initial}} = 4\pi(1+\alpha)\xi_d^3/3$, was equal to the total energy of the bubble:

$$E_{\text{bubble}} = \frac{4\pi}{3} \int_0^{\xi_d} \xi^2 \gamma^2 (3 + v^2) d\xi. \tag{15}$$

In all cases, we obtained $E_{\text{initial}} = E_{\text{bubble}}$ to within about a percent.

Although we cannot write analytic expressions for $v(\xi)$ and $w(\xi)$, analytic expressions are easily obtained in the weak-detonation limit, $\alpha \to 0$. If $\alpha \ll 1$, then the fluid velocity $v \ll 1$ everywhere, and $\xi_d - c_s \ll 1$. In this case, the small $v$ and $\xi - c_s$ expression of Steinhardt [8] can be used to describe the entire velocity profile:

$$\xi = c_s + \frac{2}{3}v\left(\ln\frac{v_d}{v} + 1\right). \tag{16}$$

In this limit, $v_d = \sqrt{3\alpha/2}$, $\xi_d = c_s + \sqrt{2\alpha/3}$, and $w_d = w_1(1 + 2\sqrt{2\alpha})$. As $\alpha \to 0$, $\gamma^2 \to 1$, and $\mu \to c_s$, so Eq. (9) can be integrated to give

$$w_0 \sim w_d \exp(-4c_s v_d) \sim w_d(1 - 2\sqrt{2\alpha}). \tag{17}$$

The enthalpy densities inside and outside the bubble are equal to lowest order in $\alpha$, $w_0 \simeq w_1$. The stress-energy integral, Eq. (1), can be also be integrated analytically. Using $d\xi/dv = (2/3)\ln(v_d/v)$, we find

$$\int_{c_s}^{\xi_d} T(\xi)\xi^2 d\xi \sim w_1 c_s^2 \int_{c_s}^{\xi_d} v^2 d\xi$$
$$= w_1 c_s^2 \int_0^{v_d} v^2 (d\xi/dv) dv = (2/27)w_1 c_s^2 v_d^3$$
$$= \sqrt{\frac{3}{2}} \frac{1}{27} w_1 \alpha^{3/2}. \tag{18}$$

Since $\xi = r/t$, the integral over $r$, Eq. (1), is Eq. (18) times $t^3$. This should be compared with the analogous result for the case of a pure vacuum bubble, Eq. (13) in Ref. [5], which in our notation is $\int T(r) r^2 dr = w_1 \alpha t^3/4$.

In the strong-detonation limit, $\alpha \to \infty$, both $\xi_d$ and $v_d$ go to unity. Simple analytic expressions for $w(\xi)$ and $v(\xi)$ cannot be found in this limit; however, we can find a simple form for the stress-tensor integral, Eq. (1), using conservation of energy. Equating $E_{\text{initial}}$ and $E_{\text{bubble}}$,

$$w_1(1+\alpha)\xi_d^3 = \int_0^{\xi_d} w(\xi)\xi^2(3+v^2)d\xi. \tag{19}$$

In a strong detonation, $w(\xi)$ and $\gamma^2$ are both strongly peaked *at* the detonation front, so the dominant contribution to the integral comes from values of $\xi$ near $\xi_d$; furthermore, near $\xi_d$, $v(\xi) \simeq 1$, so for $\alpha \gg 1$,

$$\int_0^{\xi_d} T(\xi)\xi^2 d\xi \sim w_1\alpha\xi_d^3/4 \sim \alpha w_1/4, \tag{20}$$

which smoothly matches the pure-vacuum result, Eq. (13) in Ref. [5].

For arbitrary values of $\alpha$, we can write

$$\int_0^{\xi_d} T(\xi)\xi^2 d\xi \approx \kappa(\alpha)w_1\alpha/4, \tag{21}$$

Here, $\kappa(\alpha)$ is an efficiency factor quantifying the fraction of the available vacuum energy, or latent heat, that goes into kinetic (rather than thermal) energy of the fluid. Given the weak- and strong-detonation limits for the stress-energy integral, Eq. (18) and Eq. (20), and the values at some intermediate points that we calculate numerically, we find that $\kappa(\alpha)$ can be given approximately by

$$\kappa(\alpha) = \frac{1}{1+A\alpha}\left[A\alpha + \frac{4}{27}\sqrt{\frac{3\alpha}{2}}\right], \tag{22}$$

where $A = 0.715$. The function $\kappa(\alpha)$, along with the numerically calculated value, is plotted in Fig. 5.

No signal precedes a detonation front. Therefore, except for the regions in which the bubbles have collided, the dynamics of collision of two (or more) bubbles is simply that of the sum of the individual bubbles. This is directly analogous to the case of collision of vacuum bubbles, and justifies the use of the envelope approximation for colliding detonation bubbles as explained in the following Section. We also mention that the detonation front is stable to non-spherical perturbations and therefore remains spherical as it expands [7,12]. It has also been recently postulated that although the detonation front is spherical, the fluid behind it may undergo a transition to turbulence [13]. We discuss the gravity waves that could result from the excitation of a fully developed spectrum of turbulence in the next Section.

### B. Deflagrations

In Appendix A we present a detailed discussion of the fluid dynamics of spherical deflagration. In contrast to detonations, deflagration fronts propagate at subsonic velocities and, as shown in Appendix A, are preceded by a pre-compression shock. However, unlike in the detonation case, here $T(r)$ is *not* concentrated in a thin region (unless the transition is weak and $\xi_d$ happens to be near $c_s$, which we consider unlikely), and the thin-wall approximation does not accurately describe the bubble. This makes calculating gravity waves from deflagration bubble collisions difficult. However, there are several reasons to believe that the collision of deflagration bubbles is actually a very weak source of gravity waves. Most importantly, the smaller velocities will make deflagrations a much weaker source than detonations (the fraction of vacuum energy liberated in gravity waves is proportional to $v^3$). In addition, the propagation velocity of the deflagration front is subsonic; therefore, once the pre-compression shocks collide, signals can be sent back through the bubble, and there is no reason to expect the evolution of two (or more) bubbles to resemble the sum of individual bubbles. The spherical shape of the bubble walls is likely to be disrupted shortly after the pre-compression shocks collide. Thus, in a deflagration, there will be no large concentration of kinetic energy near the bubble walls, so gravity-wave production from the collisions should be suppressed. However, we note that deflagration bubbles may be equally as effective as detonations at stirring up turbulence, which also leads to generation of gravity waves, as discussed in the next Section.

Actually, the existence of deflagration as a possible mode for a phase transition in the early Universe has recently been questioned. It has been argued that a cosmological phase transition cannot occur via deflagration because a bubble that begins expanding via deflagration rapidly becomes unstable to detonation due to the existence of hydrodynamic instabilities [7]. On the other hand, it has also been pointed out that temperature-dependence of the propagation velocity of the bubble wall could stabilize a deflagration [11], although it seems that this conclusion applies only to very weak transitions. For all of the above reasons, and especially the fact that little gravitational radiation is expected from deflagrations, we consider only detonations in the following analysis.

## III. GRAVITATIONAL RADIATION

### A. Radiation From Colliding Bubbles

As in previous work [4,5], we use the linearized-gravity approximation in Minkowski space to compute gravity-wave production. In the phase transition considered here, we expect this approximation always to be valid; see [4] for a detailed discussion. The fundamental quantity for calculating the radiation spectrum is the Fourier transform of the stress-energy tensor:

$$T_{ij}(\hat{\mathbf{k}},\omega) = \frac{1}{2\pi}\int_0^\infty dt\, e^{i\omega t}\int d^3x\, T_{ij}(\mathbf{x})e^{-i\omega\hat{\mathbf{k}}\cdot\mathbf{x}} ; \qquad (23)$$

we adopt Weinberg's unusual normalization convention [14]. We consistently ignore any pure trace pieces of the stress tensor, such as a spatially constant thermal-energy term, as they cannot contribute to the production of gravitational radiation. The source here is a number of spherical bubbles within a sample volume, each expanding at a given velocity from a given nucleation site and time. As a detonation bubble expands, its dynamics until it meets another expanding bubble are simple, described by the combustion formalism elaborated in the previous Section. Due to its spherical symmetry, a single expanding bubble produces no gravity waves. Only after bubble collisions destroy the spherical symmetry of individual bubbles is gravitational radiation emitted. In principle, the calculation of gravity waves is straightforward: once bubbles are nucleated, simply use the appropriate equations to evolve them until the phase transition is complete. For vacuum bubbles, the Klein-Gordon equation is the necessary evolution equation, while thermal bubbles require hydrodynamic equations. The stumbling block is the complexity of the bubble configurations once collisions begin. The field or fluid equations in three spatial dimensions require intensive computational resources to solve, especially considering the dynamical range in the problem: from the thickness of the bubble wall to the Hubble radius. This difficulty prompted the development of the envelope approximation [5].

In Ref. [4], the full numerical evolution for a pair of vacuum bubbles was performed and the resulting gravity-wave emission calculated. The results scale in a simple manner with the natural length and energy scales of the problem. In particular, the peak frequency of radiation is determined by the size of the bubbles at the end of their evolution, and the radiation spectrum varies with the fifth power of this length scale. The results do *not* depend on the smaller-scale structure of the scalar field which develops in the region where two bubbles collide. This scaling result suggests that the fine details of the collision region are not important to gravity-wave production, but rather that the radiation is dominated by the gross features of the evolving bubbles, namely the uncollided bubble walls. These observations prompted the envelope approximation in [5], which consists of treating the uncollided bubble walls as infinitesimally thin energy concentrations and ignoring completely the collision regions, in effect considering only the uncollided "envelope" of the expanding bubbles. This approximation turns out to be surprisingly good. In the case of two vacuum bubbles, the envelope approximation reproduces the shape and features of the gravity-wave spectrum from detailed field evolution, and its amplitude is correct to within about 10%. The numerical utility of the approximation is illustrated by a calculation involving nearly 200 vacuum bubbles nucleated in a sample volume [5], which would be impossible with full field evolution even with extensive computational resources.

As demonstrated in Section II, detonation bubbles satisfy the conditions of the envelope approximation. Specifically, the kinetic-energy density is concentrated in a thin shell near the bubble wall. In addition, the walls propagate at supersonic velocities, so anything that happens in the collision region cannot affect the expansion of the bubble in the uncollided region. On the other hand, deflagrations will not satisfy either condition. First, the energy density is not concentrated near the bubble wall; this complicates evaluation of the stress tensor, as described below. The most serious problem, however, is that the walls propagate at subsonic velocities. This means that the spherical symmetry of the bubble walls can be disrupted shortly after the pre-compression shocks collide. Since efficient gravity-wave production requires coherent motions of large energy densities, we expect the radiation production from colliding deflagration bubbles to be substantially suppressed with respect to a detonation of similar strength.

Using the envelope approximation and ignoring the bubble-collision regions, we can divide the spatial integration in Eq. (23) into regions, one surrounding each spherical bubble centered at the bubble-nucleation site $\mathbf{x}_n$. The stress tensor becomes

$$T_{ij}(\hat{\mathbf{k}}, \omega) = \frac{1}{2\pi} \int_0^\infty dt\, e^{i\omega t} \left[ \sum_{n=1}^N e^{-i\omega \hat{\mathbf{k}} \cdot \mathbf{x}_n} \int_{S_n} d\Omega \int_0^R dr\, r^2 e^{-i\omega \hat{\mathbf{k}} \cdot \mathbf{x}} T_{ij}(r, t) \right] \tag{24}$$

where $N$ is the number of bubbles, $S_n$ is the portion of the surface of bubble $n$ that remains uncollided at time $t$, and the integration variables are chosen independently around each bubble. If the bubble wall is thin, the exponential can be factored out of the radial integral, leaving the $r$-integral over the profile of the bubble stress tensor independent of the angular integral over the uncollided bubble wall.

Given the stress-energy tensor, the total energy radiated in gravity waves into a frequency interval $d\omega$ and a solid angle $d\Omega$ is [14]

$$\frac{dE}{d\omega d\Omega} = 2G\omega^2 \Lambda_{ij,lm}(\hat{\mathbf{k}}) T^*_{ij}(\mathbf{k},\omega) T_{lm}(\mathbf{k},\omega) \tag{25}$$

where $\Lambda_{ij,lm}$ is the projection tensor for gravity waves,

$$\Lambda_{ij,lm}(\hat{\mathbf{k}}) \equiv \delta_{il}\delta_{jm} - 2\hat{k}_j\hat{k}_m\delta_{il} + \tfrac{1}{2}\hat{k}_i\hat{k}_j\hat{k}_l\hat{k}_m - \tfrac{1}{2}\delta_{ij}\delta_{lm} + \tfrac{1}{2}\delta_{ij}\hat{k}_l\hat{k}_m + \tfrac{1}{2}\delta_{lm}\hat{k}_i\hat{k}_j. \tag{26}$$

Contracting with the tensor $\Lambda_{ij,lm}$ projects out the transverse-traceless piece of the source.

We model a phase transition by assuming an exponential bubble nucleation rate per unit volume [15]:

$$\Gamma = \Gamma_0 e^{\beta t}. \tag{27}$$

Note that $\beta$ here is unrelated to the velocities $\beta_1$ and $\beta_2$ defined in the combustion analysis of the previous section. This form is a reasonable *ansatz* since in general the rate will be the exponential of a characteristic nucleation action; keeping the lowest terms in a Taylor expansion around the time of the phase transition gives Eq. (27). In general, $\beta$ is expected to be of the order $4\ln(m_{\rm pl}/T)H \simeq 100H$ for a Hubble rate $H$ [16]. Bubbles are nucleated in a sample volume according to this rate. Each bubble expands at a constant velocity until all of the sample volume has been converted to the broken phase. The walls of the expanding bubbles, treated as thin shells, constitute the stress-energy tensor $T_{ij}(\mathbf{x}, t)$ in Eq. (24).

For this form for the nucleation rate, $\beta^{-1}$ is roughly the duration of the phase transition [15], and thus $\beta^{-1}v$ is roughly the mean bubble separation (*i.e.*, the bubble size at the end of the phase transition). The frequency dependence of the spectrum is set by the time scale $\beta^{-1}$, so the characteristic frequency of the radiation is $\omega \simeq \beta$. To determine the scaling of the amplitude of the radiation spectrum, we note from Eq. (22) that for a single bubble of radius $R$,

$$\int_0^R dr\, r^2 T_{ij}(r,t) = \frac{1}{3} R^3 \kappa(\alpha) \epsilon \hat{x}_i \hat{x}_j = \frac{1}{4} R^3 \kappa(\alpha) \alpha w_1 \hat{x}_i \hat{x}_j \tag{28}$$

where $\kappa(\alpha)$ is the efficiency factor introduced previously which measures the fraction of vacuum energy $\epsilon$ converted to bulk motions of the fluid. For vacuum bubbles, $\kappa = 1$ since all of the vacuum energy goes into accelerating the bubble wall. Ignoring for the moment the $e^{i\mathbf{k}\cdot\mathbf{x}}$ factors in Eq. (24), Eqs. (25), (28), and (24) imply that for a fixed number of bubbles, $N$, $dE/d\omega \propto N(R^3\kappa\epsilon)^2$. (Note that the projection tensor $\Lambda$ contracts with the unit vectors in Eq. (28) to form a dimensionless number which depends only on the geometry of the problem.) Substituting $\beta^{-1}v$ for the length scale gives

$$\frac{dE_{GW}}{d\omega} \frac{1}{E_{\rm vac}} \propto NG(R^3\kappa\epsilon)^2/(Nv^3\beta^{-3}\epsilon) \propto Gv^3\kappa^2\alpha w_1\beta^{-3} \tag{29}$$

where $E_{\text{vac}} \simeq NR^3\epsilon \simeq N\epsilon v^3\beta^{-3}$ is the total vacuum energy in the sample volume.

The neglected exponentials correspond to the usual quadrupole approximation, $e^{i\mathbf{k}\cdot\mathbf{x}} \to 1$. Since $\mathbf{k}\cdot\mathbf{x}$ scales like $v$, the quadrupole approximation will be valid for small bubble velocities, as expected. As $v$ becomes larger, the contribution of the exponentials becomes important, and the $v^3$ scaling in Eq. (29) will not hold. In fact, for the case of vacuum bubbles, $v = 1$, the quadrupole approximation overestimates the radiation spectrum by around an order of magnitude [4]. Since the quadrupole approximation scales exactly with $v^3$, the actual spectrum's amplitude will increase more slowly with $v$ than $v^3$ for larger velocities. Our numerical results show that the deviation from $v^3$ scaling begins around $v = 0.1$; see Fig. 6.

The radiation spectrum is determined by numerically evaluating the integrals in Eq. (24) for the source configuration of many bubbles nucleated in a sample volume. We use trials with 20 to 30 bubbles because this number is computationally tractable and because significantly more bubbles give essentially the same results for the radiation efficiency, as demonstrated in Ref. [5]. Thus, for a given value of $\beta$, the physical sample volume is proportional to $v^3$, insuring that approximately the same number of bubbles will be nucleated in the sample volume for any velocity. We have five trial nucleations in a spherical sample volume, each with between 17 and 33 bubbles, nucleated randomly according to Eq. (27). These are the same nucleation trials used in Ref. [5]. We use the same nucleation trials for all bubble-expansion velocities by re-scaling all distances in the $v = 1$ case by a factor of $v$; using the same nucleation trials minimizes any spectrum differences arising simply from geometry of the bubbles. For each trial nucleation and bubble expansion velocity, we calculate the radiation-energy spectrum in the six directions $(\pm\hat{\mathbf{x}}, \pm\hat{\mathbf{y}}, \pm\hat{\mathbf{z}})$, and then average over the five trials and six directions to obtain a mean spectrum. These spectra are plotted as power per octave for various velocities in Fig. 7. The statistical variation in the mean due to the averaging is around 10%. Each spectrum peaks at a characteristic frequency of around $2\beta$ independent of bubble expansion velocity, as expected. In Fig. 6, we plot the ratio of energy radiated in gravity waves to the total energy (thermal plus vacuum energy); the straight line displays $v^3$ scaling. The departure from $v^3$ scaling as $v \to 1$ is clear. The solid curve is the analytic fit to the fraction of energy liberated into gravity waves,

$$\frac{E_{GW}}{E_{\text{tot}}} \approx 0.07\kappa^2 \left(\frac{H}{\beta}\right)^2 \left(\frac{\alpha}{1+\alpha}\right)^2 \left(\frac{v^3}{0.24 + v^3}\right). \qquad (30)$$

Note that in the strong-detonation limit, $v \to 1$ and $\alpha \to \infty$, this reduces to the vacuum-bubble result of Ref. [3].

The radiation spectra in Fig. 7 depend on the parameters $v$, $\kappa$, $\beta$, and $\epsilon = 3w_1\alpha/4$. A particular phase transition is characterized by the temperature at which it occurs and its latent heat, or equivalently by $w_1$ and $\alpha$. For detonation bubbles, $v$ and $\alpha$ are related by Eq. (13), and $\kappa$ and $\alpha$ by Eq. (22). The parameter $\beta$ describing the bubble-nucleation rate will be determined by the effective action for nucleating bubbles. Thus we have assembled all the necessary ingredients to calculate the gravity waves produced by a thermal first-order phase transition which proceeds via detonation bubbles.

## B. Radiation From Fully Developed Turbulence

Injection of energy into the universe will cause turbulence if the Reynolds number of the early-Universe plasma is large enough at the time of energy injection. Here we estimate the gravity waves produced by a Kolmogoroff spectrum of turbulence, independent of any details of the phase transition dynamics.

The Reynolds number in the early-Universe plasma is very large for length scales $L$ not too different than the Hubble radius $H^{-1} \sim m_{\rm Pl}/T^2$. Specifically, the Reynolds number $Re = LV/\nu \simeq \gamma g^4 (m_{\rm Pl}/T)$, with $L \equiv \gamma H^{-1}$, the kinematic viscosity $\nu \simeq v\ell$, $\ell \simeq 1/n\sigma \simeq 1/g^4 T$ is the particle mean-free path ($g$ is a typical gauge coupling and $T$ is the plasma temperature), and $V/v =$ (bulk flow velocity)/(microscopic velocity) is taken to be of order unity. Thus, it is quite reasonable to expect turbulence to develop when the plasma is "stirred up" by a phase transition (the critical Reynolds number for the onset of turbulence is around 2000), especially if bubble walls are unstable to perturbations and become highly nonspherical.

In the case of fully developed turbulence the distribution of the turbulent kinetic-energy density is expected to take the stationary Kolmogoroff form [17],

$$k \frac{d\rho_{\rm turb}}{dk} \propto k^{-2/3}, \tag{31}$$

which is characterized by a constant flow of turbulent kinetic energy from larger scales to smaller scales,

$$\frac{\rho v_L^2}{\tau_L} = \frac{k}{\tau_L} \frac{d\rho_{\rm turb}}{dk} = {\rm const}; \tag{32}$$

here $\rho$ is the plasma energy density. The turbulent velocity associated with an eddy of size $L \simeq k^{-1}$, $v_L$, and its lifetime, $\tau_L$, are related, $\tau_L \simeq L/v_L$. For the Kolmogoroff spectrum

$$v_L \propto L^{1/3}; \qquad \tau_L \propto L^{2/3}. \tag{33}$$

That is, an eddy survives for about a turnover time before it breaks into smaller eddies. (So long as the eddy survival time is a scale-independent factor times the eddy turnover time, the Kolmogoroff spectrum should develop.)

On very small scales, $k \gtrsim k_D$, the spectrum is cutoff due to viscous damping of eddies. The damping scale $k_D$ is the scale on which viscosity diffuses the turbulence as fast as the transfer of kinetic energy from larger scales replenishes it: $\tau_{\rm dif} \simeq L^2/\ell \simeq \tau_L$; for the Kolmogoroff spectrum $k_D \propto \ell^{-3/4}$. On scales $k \gg k_D$, $k d\rho_{\rm turb}/dk \propto k^{-6}$.

The Kolmogoroff spectrum is established as turbulence is introduced on some large scale—*e.g.*, by the "stirring" of the plasma by expanding bubbles—and is fed down to small scales as large eddies break into smaller eddies. It takes of the order of an eddy turnover time on the largest length scale to establish the Kolmogoroff spectrum. The stationary spectrum of turbulence persists as long as the plasma is being stirred. Once the stirring stops, the turbulence dissipates in about a turnover time for the largest length scale.

Next, let us estimate the amount of gravitational radiation produced by eddies of characteristic size $L$. Using the quadrupole formula, $P_{\rm GW} \simeq G(d^3Q/dt^3)^2$, and estimating the triple time derivative of the quadrupole moment of a typical eddy as $d^3Q/dt^3 \simeq L^3\rho v_L^2/\tau_L$, it follows that the volume density of gravitational radiation produced by eddies of size $L$ is

$$\omega \frac{d\rho_{\rm GW}}{d\omega} \simeq G\rho^2 L^3 \mathcal{T} v_L^4/\tau_L^2 \propto \omega^{-9/2} \tag{34}$$

where time $\mathcal{T}$ is the duration of the turbulence and the characteristic frequency $\omega \simeq \tau_L^{-1} \simeq v_L/L \simeq v_L k$. In making this estimate we have made two reasonable assumptions: (i) that the quadrupole moment of an eddy varies by order unity on a turnover time; and (ii) that the radiation from different eddies adds incoherently. Like the turbulent kinetic energy itself, the energy in gravitational radiation achieves its maximum on the largest length scale.

Finally, let us be more specific. Suppose that the largest length scale on which the turbulence is being driven is $L_0 \equiv \beta^{-1} v$, and that the fluid velocities on this length scale are $v_0$ (*not* to be confused with the velocity $v$ of propagation of the bubble wall). Further, we assume that the turbulence persists for a time $\mathcal{T} \approx \beta^{-1}$, corresponding to the length of the phase transition. Then we have the following approximate relations:

$$v_L \simeq \left(\frac{L}{L_0}\right)^{1/3} v_0; \quad \tau_L \simeq \frac{L}{v_L} \simeq L^{2/3} L_0^{1/3} v^{-1}; \tag{35}$$

$$k_D \simeq (vL_0/\ell)^{3/4} L_0^{-1} \simeq L_D^{-1}; \quad \omega_D \simeq \tau_D^{-1} \simeq v k_D. \tag{36}$$

It then follows that the spectrum of the energy density in gravity waves is

$$\frac{\omega}{\rho} \frac{d\rho_{\rm GW}}{d\omega} \simeq \left(\frac{H}{\beta}\right)^2 v v_0^6 \left(\frac{\omega}{\omega_0}\right)^{-9/2} \tag{37}$$

$$\omega_0 \simeq \tau_{L_0}^{-1} \simeq \beta v^{-1} v_0, \tag{38}$$

where this spectrum extends from frequency $\omega_0$ up to $\omega_D$.

Strictly speaking, these expressions are valid only in the regime of nonrelativistic fluid velocities, $v_0 \ll 1$, and likely overestimate the gravity-wave production if applied to a stronger transition. For a detonation, the initial fluid velocity $v_0$ can be estimated from the fraction of the total energy that goes into kinetic energy of the fluid. Thus, in the weak-detonation limit, $v_0 \sim (\kappa\alpha)^{1/2}$, and in the strong-detonation limit, $v_0 \sim 1$. For a deflagration, the fluid velocity may be estimated by Eq. (A1).

Our estimate for the gravitational radiation produced in a phase transition should be viewed as an absolute, albeit approximate, lower bound. No account was made of the radiation emitted by the bubble walls themselves; only that arising from the turbulent motion of the plasma that was stirred up by the release of the latent heat was taken into account. Further, we wish to emphasize that our analysis and estimates should apply to any violent injection of energy on large scales in the early Universe.

## IV. RELIC GRAVITY WAVES

To translate the results of the previous section into the potentially observable background of gravity waves today, we must propagate the gravity waves forward from the phase transition until today. This is simple since the gravity waves are essentially decoupled from the rest of the universe. The energy density in gravity waves decreases as $R^{-4}$, and the frequency of the gravity waves redshifts as $R^{-1}$, where $R$ is the scale factor. If the universe has expanded adiabatically since the phase transition, meaning that the entropy per comoving volume $S \propto R^3 g(T) T^3$ remains constant, then the ratio of the scale factor at the transition to the scale factor today is given by

$$\frac{R_*}{R_0} = 8.0 \times 10^{-14} \left(\frac{100}{g_*}\right)^{1/3} \left(\frac{1\,\mathrm{GeV}}{T_*}\right). \tag{39}$$

In these expressions, $g(T)$ counts the total number of relativistic degrees of freedom at a given temperature, and the star subscript refers to the value of a quantity at the time of the phase transition. If we denote the fraction of total energy density in gravity waves at the transition as $\Omega_{GW*}$ and the characteristic frequency at the transition as $f_*$, then the fraction of critical density today $\Omega_{GW}$ and characteristic frequency $f_0$ today are

$$f_0 = f_* \left(\frac{R_*}{R_0}\right) = 1.65 \times 10^{-7}\,\mathrm{Hz} \left(\frac{f_*}{H_*}\right) \left(\frac{T_*}{1\,\mathrm{GeV}}\right) \left(\frac{g_*}{100}\right)^{1/6} \tag{40}$$

$$\Omega_{GW} = \Omega_{GW*} \left(\frac{R_*}{R_0}\right)^4 \left(\frac{H_*}{H_0}\right)^2 = 1.67 \times 10^{-5} h^{-2} \left(\frac{100}{g_*}\right)^{1/3} \Omega_{GW*}, \tag{41}$$

where $h$ is the current value of the Hubble parameter in units of $100\,\mathrm{km\,sec^{-1}\,Mpc^{-1}}$, and we have used the relation

$$H_*^2 = \frac{8\pi G \rho_{\mathrm{rad}}}{3} = \frac{8\pi^3 g_* T_*^4}{90 m_{\mathrm{Pl}}^2}. \tag{42}$$

We also define a characteristic amplitude $h_c(f)$ produced by stochastic gravity waves around frequency $f$ as

$$h_c(f) \equiv 1.3 \times 10^{-18} [\Omega_{GW}(f) h^2]^{1/2} \left(\frac{1\,\mathrm{Hz}}{f}\right), \tag{43}$$

where $\Omega_{GW}(f)$ is the contribution per frequency octave to the energy density in gravity waves [18].

Using the results in the previous section, we can describe the gravity waves from bubble collisions by

$$\Omega_{GW} h^2 \approx 1.1 \times 10^{-6} \kappa^2 \left(\frac{H_*}{\beta}\right)^2 \left(\frac{\alpha}{1+\alpha}\right)^2 \left(\frac{v^3}{0.24 + v^3}\right) \left(\frac{100}{g_*}\right)^{1/3} \tag{44}$$

$$f_{\max} \approx 5.2 \times 10^{-8}\,\text{Hz}\left(\frac{\beta}{H_*}\right)\left(\frac{T_*}{1\,\text{GeV}}\right)\left(\frac{g_*}{100}\right)^{1/6},\tag{45}$$

$$h_c(f_{\max}) \approx 1.8 \times 10^{-14}\kappa\left(\frac{\alpha}{1+\alpha}\right)\left(\frac{H_*}{\beta}\right)^2\left(\frac{1\,\text{GeV}}{T_*}\right)\left(\frac{v^3}{0.24+v^3}\right)^{1/2}\left(\frac{100}{g_*}\right)^{1/3}.\tag{46}$$

For detonation bubbles, in the weak-transition limit $\alpha \to 0$, $\kappa \propto \sqrt{\alpha}$, so the amplitude of gravity waves is suppressed by a factor of $\alpha^{3/2}$ relative to the amplitude in the case of a pure-vacuum transition.

For the case of turbulent mixing, the same analysis applies though our estimates are much rougher. We assume that after the phase transition the ratio of the energy density in gravitational waves to that in radiation is of the order of $\Omega_{GW*} \simeq (H_*/\beta)^2 v\alpha^3\kappa^3$ and the spectrum peaks at the frequency $2\pi f_* \simeq \beta v^{-1}\alpha^{1/2}\kappa^{1/2}$. Then we have the following estimates:

$$\Omega_{GW}h^2 \simeq 10^{-5}\left(\frac{H_*}{\beta}\right)^2 vv_0^6\left(\frac{100}{g_*}\right)^{1/3},\tag{47}$$

$$f_{\max} \simeq 2.6 \times 10^{-8}\,\text{Hz}\,v_0 v^{-1}\left(\frac{\beta}{H_*}\right)\left(\frac{T_*}{1\,\text{GeV}}\right)\left(\frac{g_*}{100}\right)^{1/6},\tag{48}$$

$$h_c(f_{\max}) \simeq 5 \times 10^{-13}\,v_0^2\left(\frac{H_*}{\beta}\right)^2\left(\frac{1\,\text{GeV}}{T_*}\right)\left(\frac{100}{g_*}\right)^{1/3}.\tag{49}$$

Note that the characteristic amplitude for gravity waves from bubble collisions and from turbulence scales in the same way, and our rough estimates indicate that fully-developed turbulence is comparable to, and maybe more potent than, bubble collisions in generating gravity waves.

For a particular first-order phase transition, knowledge of the parameters $v$, $\beta$, $\kappa$, and $\alpha$ suffice to determine the resulting gravity-wave spectrum from bubble collisions. For detonation bubbles, $v$ and $\kappa$ are functions of $\alpha$ (cf. Figs. (1) and (5). In contrast the time scale $\beta$ and the energy scale $\alpha$ are determined entirely by the bubble-nucleation probability. In terms of fundamental physical quantities, $\beta$ and $\alpha$ are determined by the effective potential for bubble nucleation. Knowledge of the mean bubble separation $L_0 = \beta^{-1}v$ and the characteristic fluid velocity $v_0$ suffice to determine the spectrum of gravitational radiation from turbulence resulting from the transition.

As a direct application of our general formalism, we consider the electroweak phase transition. This cosmological phase transition has been the focus of much attention recently. If the electroweak phase transition was first order, then the baryon asymmetry of the Universe may have been produced at the electroweak phase transition [19]. Such a transition would have produced gravitational radiation; we now use our results to estimate the strength of this signal.

The minimal standard model electroweak phase transition occurs when the $SU(2)_L \times U(1)_Y$ gauge symmetry is broken to $U(1)_{EM}$. The bubble-nucleation rate and latent heat of the transition follow from the effective potential for the Higgs field $\phi$. In Appendix B, we review a general form for the effective potential and its specific realization for a one-loop electroweak calculation. We adopt the reference values $m_t = 100\,\text{GeV}$ for the top mass and $m_H = 60\,\text{GeV}$ for the Higgs mass; the end of Appendix B shows how the relevant parameters vary with these masses. The transition then occurs at a temperature $T_* \approx 104\,\text{GeV}$ and results in $H_*/\beta = 1.3 \times 10^{-3}$, $\alpha = 1.4 \times 10^{-3}$, $\kappa = 7.8 \times 10^{-3}$, and $v = c_s = 0.57$. Then for bubble collisions, we get $\Omega h^2 \approx 9.8 \times 10^{-23}$ and $h_c \approx 1.5 \times 10^{-27}$, peaking at a frequency around $f_{\max} \approx 4.1 \times 10^{-3}\,\text{Hz}$. Reasonable changes in the reference values for the Higgs and top masses and uncertainties in the accuracy of the one-loop effective potential could conceivably change these values by an order of magnitude or more. The weak gravity-wave signal that results from the electroweak phase transition is a consequence of the fact that the transition in the standard model is very weakly first order, if first order at all.

Various generalizations of the standard model, particularly enlarged Higgs sectors in supersymmetric models, can substantially strengthen the electroweak transition [20]. Other more speculative first-order transitions, such as in various GUT theories, may also have taken place. We can ask what characteristics must a first-order phase transition possess to generate a gravity wave signal which is potentially detectable. For the LIGO facility with advanced detectors, the ultimate sensitivity to a stochastic background is an amplitude of around $2 \times 10^{-25}$ at 100 Hz [18,21]. Requiring the peak frequency of the radiation spectrum to fall at 100 Hz, the most sensitive LIGO frequency, gives $(\beta/H_*)(T_*/1\,\text{GeV}) \simeq 2 \times 10^9$ by Eq. (45). Then for the expected value of $\beta/H_* \simeq 100$, Eq. (46) gives $h_c \simeq 9 \times 10^{-26} \kappa\alpha/(1+\alpha)$ at the peak frequency, making detection by LIGO marginal at best.

The situation is more promising for a space-based interferometer. Projected capabilities of a long baseline interferometer between two satellites are a frequency range from $10^{-5}$ to $10^{-1}$ Hz, and a sensitivity down to an amplitude of $10^{-22}$ at $10^{-4}$ Hz [18,22]. In this case, requiring the peak of the gravity wave spectrum to fall at $10^{-4}$ Hz gives $(\beta/H_*)(T_*/1\,\text{GeV}) \simeq 2 \times 10^3$. Again taking $\beta/H_* \simeq 100$, this corresponds to a phase transition temperature of 20 GeV; the characteristic amplitude of the gravity waves is $h_c \simeq 10^{-19}\kappa\alpha/(1+\alpha)$. This background is detectable as long as $\kappa\alpha/(1+\alpha) \gtrsim 10^{-3}$, a reasonable condition for a strong phase transition. These estimates can be made less stringent by noting that the gravity wave spectrum for colliding bubbles falls slowly with frequency, and that measuring the gravity wave background at a frequency 10 or 100 times higher than the peak frequency only results in the amplitude dropping by a factor of a few. We have also not included any gravity waves from turbulence, which could give a comparable and independent contribution. A strong electroweak phase transition at $T = 100\,\text{GeV}$ is potentially detectable by a space-based interferometer.

In conclusion, we have calculated the gravitational radiation produced by two potentially strong sources during a first-order phase transition: the collision of spherically symmetric bubbles, and fully-developed turbulence. Detailed numerical simulation of many colliding bubbles leads to a characteristic radiation spectrum which scales with $\alpha$, $\kappa$, and $\beta$, parameters related to the latent heat, efficiency, and time scale of the transition respectively; the spectrum also depends on the bubble expansion velocity $v$ in a sensible way. Relativistic detonation bubbles provide a simple model for bubble dynamics which allows $\kappa$ and $v$ to be expressed in terms of $\alpha$. Likewise, estimates of the radiation spectrum from stationary Kolmogoroff turbulence give similar scalings with these parameters. These estimates indicate that turbulence is likely as potent a source of gravitational radiation as bubble collisions. The magnitude of the frequency and amplitude of the resulting gravity-wave stochastic background makes detection of a strong phase transition by a future space-based interferometer an open possibility, but makes unlikely detection of a first-order phase transition by the upcoming LIGO detectors.

## ACKNOWLEDGMENTS

We thank Dave DeYoung for several helpful discussions about turbulence. MK has been supported in part by the Texas National Research Laboratory Commission and by DOE through Grant No. DE-FG02-90ER40542. AK and MST are supported in part by the DOE (at Chicago and Fermilab) and by NASA through grant No. NAGW 2381 (at Fermilab). AK is supported in part by the NASA Graduate Student Researchers Program.

## APPENDIX A: FLUID FLOW IN DEFLAGRATIONS

Here we present a detailed analysis of deflagration bubbles, analogous to that of detonations in Sec. II.A. Our aim is to determine the radial-velocity profile of the deflagration bubble.

We again start with Eq. (7). If we are considering deflagrations, then in the wall frame, fluid flows into the discontinuity with a velocity $v_1$ and out of the wall frame with a velocity $v_2 > v_1$, and both $v_2, v_1 < c_s$. In the case of spherical deflagration, since the fluid at the center of the bubble is at rest, this means that (in the "laboratory" frame) the wall propagates at a velocity $v_2$, so the fluid velocity is $v = 0$ for $\xi < v_2$. Since $v_2 > v_1$, the expansion of the gas during combustion exerts a piston effect on the fluid outside the bubble and pushes the fluid just outside the bubble with a velocity

$$v(\xi = v_2) = \frac{v_2 - v_1}{1 - v_1 v_2} \equiv v_0. \tag{A1}$$

So in order to determine the radial velocity profile in a spherical deflagration, we need to solve Eq. (7) subject to the boundary condition Eq. (A1). This is straightforward.

To begin, note that since $v, \xi, (1 - v\xi), \gamma^2 > 0$ always, $dv/d\xi < 0$ as long as $\mu < c_s$. Since $\mu \leq c_s$ for $\xi \leq c_s$ (the equalities holding only if $v = 0$ and $\xi = c_s$), we know that $dv/d\xi < 0$ and that $v$ is always decreasing for $\xi < c_s$. The fluid far from the center of the bubble is at rest, so for some value of $\xi < 1$, the fluid velocity $v$ goes to zero. The question is whether this occurs for (i) $\xi < c_s$, (ii) $\xi = c_s$, or (iii) $\xi > c_s$.

If at some value of $\xi$, $v \to 0$, then $\ln v \to -\infty$, and $d(\ln v)/d\xi \to -\infty$; however, $d(\ln v)/d\xi \to -\infty$ if and only if the quantity in brackets in the left-hand side of Eq. (7) goes to zero (i.e. $\mu = c_s$). Since this does not occur for $\xi < c_s$, the fluid velocity $v$ does *not* decrease to zero for $\xi < c_s$.

Now if we suppose that $v \to 0$ at $\xi = c_s$, then we can study Eq. (7) in the limit $v \ll 1$, $(\xi - c_s) \ll 1$, and we find that the solution in this case is [8]

$$\xi - c_s = \frac{2}{3} v \ln \frac{v_0}{v}, \qquad v, \xi - c_s \ll 1. \tag{A2}$$

For $v > 0$ the right-hand side is always positive, but the left-hand side is negative for $\xi < c_s$, so there is *no* solution to Eq. (7) where the velocity goes to zero at $\xi = c_s$.

Therefore, the radial velocity must go to zero for some value of $\xi > c_s$. Again, if $v$ is to go smoothly to zero, then $d \ln v/d\xi \to -\infty$, as $v \to 0$. It is clear from Eq. (7) that this cannot occur for $\xi > c_s$, so a discontinuity must occur, and as we may have guessed for supersonic propagation, there must be a shock. Although $d \ln v/d\xi$ does not diverge as $v \to 0$, it does go to $-\infty$ for some $\xi > c_s$; this occurs when $\mu = c_s$ [where $v = (\xi - c_s)/(1 - v\xi)$]. So assume that this is where the physical discontinuity occurs. Doing so, we find that in the frame of the discontinuity, fluid flows into the discontinuity with a velocity $\beta_1 = \xi$ and flows out of the discontinuity with a velocity $\beta_2 = c_s (\neq 1/3\beta_1)$. In a shock, $\beta_1 = 1/3\beta_2$ [8], so this discontinuity *cannot* be physical. Therefore, the shock must occur at some value of $\xi$ less than that at which $\mu = c_s$.

To find the value of $\xi$ at which the shock occurs, we again note that in the frame of the discontinuity the velocities of the fluid in and out of the discontinuity are $\beta_1 = \xi$ and $\beta_2 = \mu$, and then note that in a shock $\beta_1 = 1/3\beta_2$. This then tells us that the shock occurs when

$$\frac{\xi}{c_s} \frac{\mu}{c_s} = 1. \tag{A3}$$

It is reassuring to note that this occurs for a value of $\xi$ smaller than that at which $d \ln v/d\xi$ diverges (determined by $\mu/c_s = 1$).

So, to determine the velocity profile (and from it the stress-energy tensor) for a spherical deflagration bubble, Eq. (7) is integrated subject to the boundary condition, Eq. (A1), until $\mu\xi/c_s^2 = 1$. At this point there is a shock. As the strength of the transition is increased, $v_0$ will increase, and the value of $\xi$ at which the shock occurs will increase. This simply means that the strength of the pre-compression shock preceding the deflagration front increases as the strength of the transition increases.

Generally, Eq. (7) must be solved numerically, but if the transition is weak, then $v_2 \simeq v_1$ and $v_0 \ll 1$. In the limit of small velocities ($v \ll 1$, and as long as $\xi - c_s$ is not too small), Eq. (7) becomes

$$\left(\frac{\xi^2}{c_s^2} - 1\right) \frac{dv}{d\xi} = \frac{2v}{\xi}, \tag{A4}$$

which can be integrated subject to the boundary condition $v(\xi_0) = v_0$, to give

$$v(\xi) = v_0 \left(\frac{\xi_0}{\xi}\right)^2 \frac{c_s^2 - \xi^2}{c_s^2 - \xi_0^2}. \tag{A5}$$

According to this solution, near the deflagration front, the radial velocity falls off quadratically with radius and then begins to decrease even faster and goes to zero at $\xi = c_s$. Strictly speaking, this solution is not valid at $\xi = c_s$ and the radial velocity does *not* go to zero exactly at $\xi = c_s$, but if the transition is indeed weak, the pre-compression shock will be at a value of $\xi$ just slightly larger than $\xi = c_s$, and Eq. (A5) should provide a good approximation to $v(\xi)$. In Fig. 8, we plot the fluid velocity as a function of $\xi$ for a rather weak deflagration ($v_2 = 0.1$ and $v_0 = 0.01$). We plot the fluid velocity as function of $\xi$ for stronger deflagrations in Fig. 9; the dashed curve illustrates a deflagrations with $v_2 = 0.1$ and $v_0 = 0.09$, and the solid curve illustrates the case where $v_2 = 0.5$ and $v_0 = 0.45$.

The fluid flow in a spherical deflagration is different from that in a planar deflagration [9–11]. In a planar deflagration, the velocity of the fluid between the deflagration front and the pre-compression shock is constant. On the other hand, the fluid velocity and enthalpy density decrease with increasing $\xi$ in spherical deflagration, as we have shown. Therefore, for given values of $\beta_1$ and $\beta_2$, the pre-compression shock is weaker in a spherical deflagration than it would be in a planar deflagration, and in the limit of a weak transition, it is much weaker. (Similar conclusions were obtained for non-relativistic deflagrations [23]). Consequently, the allowable modes of deflagration in a phase transition in the early Universe may be slightly different than those discussed previously [9–11].

### APPENDIX B: THE EFFECTIVE POTENTIAL FOR BUBBLE NUCLEATION

Calculation of the gravity waves from a first-order phase transition requires two essential pieces of information about the transition: the parameters $\alpha$ and $\beta$. These parameters characterize the overall properties of the transition and follow from the effective potential for bubble nucleation.

#### 1. A Model Effective Potential

In a typical first-order phase transition, the probability for nucleation of a low-temperature phase bubble will be determined by the tunneling action between two vacua of an effective potential. To parameterize this effective potential, we consider the general form

$$V(\phi, T) = \frac{1}{2}\gamma(T^2 - T_0^2)\phi^2 - \frac{1}{3}\alpha T \phi^3 + \frac{1}{4}\lambda_T \phi^4 \tag{B1}$$

where $\gamma$, $\alpha$, and $\lambda$ are arbitrary positive constants and $T_0$ sets the temperature scale [10,24]. This potential possesses two inequivalent minima. The symmetric phase potential minimum is always at $\phi = 0$ where $V(\phi) = 0$. The broken phase minimum occurs at

$$\phi = v(T) \equiv \frac{\alpha T}{2\lambda}\left(1 + \sqrt{1 - \frac{8x}{9}}\right) \tag{B2}$$

where we have defined

$$x \equiv \frac{9\gamma\lambda}{2\alpha^2}\frac{(T^2 - T_0^2)}{T^2} = \frac{t_0 - t}{t_0 - t_c}. \tag{B3}$$

In the second expression for $x$, we have presumed a quadratic relation between time and temperature, $t_c T_c^2 = t_0 T_0^2$, valid in a radiation dominated universe at constant entropy. The critical temperature $T_c$ at which the free energy of the symmetric and broken phases are equal is given by the relation

$$T_c^2\left(1 - \frac{2\alpha^2}{9\gamma\lambda_T}\right) = T_0^2. \tag{B4}$$

At the critical temperature, the energy density of the broken phase first dips below that of the symmetric phase; at the temperature $T_0$, the symmetric phase becomes unstable. A first-order phase transition occurs at a temperature $T_*$, with $T_c > T_* > T_0$.

To determine the latent heat and vacuum energy associated with the transition, we begin with the value of the potential at the broken phase minimum:

$$B(T) = -V(v(T), T) = \frac{\alpha^4 T^4}{24\lambda^3} \left[ \frac{8x^2}{27} - \frac{4x}{3} + 1 + \left(1 - \frac{8x}{9}\right)^{3/2} \right], \tag{B5}$$

which is the difference in free energy density between the two states of the system. The derivative of $B$ is given by

$$\frac{dB}{dT} = v^2(T) \left( -\gamma T + \frac{\alpha v(T)}{3} \right) \tag{B6}$$

and latent heat is defined as

$$L \equiv -T_c \frac{dB}{dT} \bigg|_{T_c} = \frac{4\alpha^2 \gamma}{9\lambda^2} T_0^2 T_c^2. \tag{B7}$$

The vacuum energy associated with the transition is [10]

$$\epsilon = B(T) - T B'(T). \tag{B8}$$

To calculate $\beta$ for a given phase transition, the basic quantity we need is $\Gamma(t) = A e^{-S(t)}$, the bubble-nucleation rate per unit volume per unit time. The dimensionful prefactor $A$ is expected to be of order $T_c^4$ but is unimportant for the present calculation. The argument in the exponential is the action for nucleating critical bubbles. At high temperatures, this action is well-approximated by [25]

$$S(t) = \frac{123 \gamma^{3/2}}{\alpha^2} \left( \frac{T - T_0}{T} \right)^{3/2} F\left( \frac{9\lambda_0 \gamma (T - T_0)}{\alpha^2 T} \right)$$
$$= 13.7 \alpha \lambda^{-3/2} x^{3/2} F(x) \tag{B9}$$

where the function $F$ is defined by

$$F(x) \equiv 1 + \frac{x}{4} \left[ 1 + \frac{2.4}{1-x} + \frac{0.26}{(1-x)^2} \right]. \tag{B10}$$

This parameterization is accurate to around 1% for $0 < x < 0.95$ [25].

The nucleation rate is a rapidly increasing function of time near the phase transition, so it is sensible to expand the action in a Taylor series about $t = t_*$: [15]

$$S(t) \approx S_* - \beta(t - t_*), \tag{B11}$$

$$\beta = -\frac{dS}{dt} \bigg|_{t=t_*} = -\frac{9\lambda\gamma}{2\alpha^2} \frac{1}{tS} \frac{dS}{dx} \bigg|_{t=t_*} > 0. \tag{B12}$$

Then the nucleation rate can be rewritten as $\Gamma = \Gamma_0 \exp \beta t$ as in the previous Section. Simple estimates show that the electroweak transition takes place when $S \approx 130$ [10,25].

## 2. The Electroweak Case

The exact parameters of the electroweak symmetry breaking phase transition are not yet well known, due both to uncertainties in the standard model (*e.g.*, the top and Higgs masses) and to theoretical difficulties in calculating the effective potential, which determines the order of the phase transition and the bubble-nucleation rate. For the present calculation, we use the one-loop approximation to the finite-temperature effective potential [26] with an improved cubic term [25]: where the coefficients are given by

$$\gamma = \frac{1}{4v_0^2}(2m_W^2 + m_Z^2 + 2m_t^2), \tag{B13a}$$

$$\alpha = \frac{1}{2\pi v_0^3}(2m_W^3 + m_Z^3), \tag{B13b}$$

$$T_0^2 = \frac{1}{2\gamma}\left[m_H^2 - \frac{3}{8\pi^2 v_0^2}(2m_W^4 + m_Z^4 - 4m_t^4)\right], \tag{B13c}$$

$$\lambda_T = \frac{m_H^2}{2v_0^2} - \frac{3}{16\pi^2 v_0^4}\left(2m_W^4 \ln\frac{m_W^2}{a_B T^2} + m_Z^4 \ln\frac{m_Z^2}{a_B T^2} - 4m_t^4 \ln\frac{m_t^2}{a_F T^2}\right) \tag{B13d}$$

with $v_0 = 246\,\text{GeV}$, $\ln a_B \approx 3.51$, and $\ln a_F \approx 1.14$.

We adopt the following reference values: W mass $m_W = 80.6\,\text{GeV}$, Z mass $m_Z = 91.2\,\text{GeV}$, top mass $m_t = 100\,\text{GeV}$, and Higgs mass $m_H = 60\,\text{GeV}$. With these masses, the above coefficients have the values $\gamma = 0.17$, $\alpha = 0.019$, $T_0 = 103.6\,\text{GeV}$, and $\lambda_0 \equiv \lambda_T(T = T_0) = 0.028$. The Higgs self-coupling $\lambda_T$ depends very weakly on $T$, and we will ignore the variation in $\lambda_T$ over the temperature range of interest.

For the above parameters, $x = 0.74$ if the phase transition occurs when $S = 130$. Then Eq. (B8) gives $\epsilon = 0.049 T_*^4$ so

$$\alpha = 30\epsilon/\pi^2 g_* T_*^4 = 1.4 \times 10^{-3}. \tag{B14}$$

Working out the derivative in Eq. (B12) leads to $\beta \simeq 400/t_*$, which gives

$$\frac{H_*}{\beta} = 1.3 \times 10^{-3}, \tag{B15}$$

using the relationship $t_* = 0.30 m_{\text{Pl}}/T_*^2 g_*^{1/2}$. Since $\alpha$ is so small, Eq. (13) shows that the expansion velocity of detonation bubbles is essentially $v = c_s = 1/\sqrt{3}$. Finally, the fraction of the vacuum energy which goes into bubble wall kinetic energy is, by Eq. (22), $\kappa = 7.8 \times 10^{-3}$.

# REFERENCES

[1] M. S. Turner and F. Wilczek, *Phys. Rev. Lett.* **65**, 3080 (1990).
[2] C. J. Hogan, *Mon. Not. R. Astron. Soc.* **218**, 629 (1986); E. Witten, *Phys. Rev. D* **30**, 272 (1984).
[3] A. Kosowsky, M. S. Turner, and R. Watkins, *Phys. Rev. Lett.* **69**, 2026 (1992).
[4] A. Kosowsky, M. S. Turner, and R. Watkins, *Phys. Rev. D* **45**, 4514 (1992).
[5] A. Kosowsky and M. S. Turner, *Phys. Rev. D* **47**, 4372 (1993).
[6] S. Coleman, *Phys. Rev. D* **15**, 2929 (1977); C. G. Callan and S. Coleman, *ibid.* **16**, 1762 (1977).
[7] M. Kamionkowski and K. Freese, *Phys. Rev. Lett.* **69**, 2743 (1992).
[8] P. J. Steinhardt, *Phys. Rev. D* **25**, 2082 (1982).
[9] M. Gyulassy, K. Kajantie, H. Kurki-Suonio, and L. McLerran, *Nucl. Phys.* **B237**, 477 (1984); L. van Hove, *Z. Phys. C*, **21**, 93 (1983).
[10] K. Enqvist, J. Ignatius, K. Kajantie, and K. Rummukainen, *Phys. Rev. D* **45**, 3415 (1992).
[11] P. Huet *et al.*, *Phys. Rev. D* **48**, 2477 (1993).
[12] L. Landau and D. Lifshitz, *Fluid Mechanics* (Pergamon, New York, 1959).
[13] M. Abney, FERMILAB-PUB-93/098-A (1993).
[14] S. Weinberg, *Gravitation and Cosmology* (Wiley, New York, 1972), Chap. 10.
[15] M. S. Turner, E. J. Weinberg, and L. M. Widrow, *Phys. Rev. D* **46**, 2384 (1992).
[16] The exceptions to the expectation that $\beta \simeq 100H$ are first-order inflationary phase transitions. For a discussion of the exponential nucleation rate approximation and expectations for $\beta$ see Ref. [15]. (Note, the discussion in Ref. [3] is derived from Ref. bubnuc, where a more thorough treatment is given.)
[17] A.N. Kolmogoroff, *Dokl. Akad. Nauk. SSSR* **30**, 299 (1941).
[18] K. S. Thorne, in *300 Years of Gravitation*, ed. S. Hawking and W. Israel (Cambridge Univ. Press, Cambridge, 1987).
[19] See, *e.g.*, A. Cohen, D. Kaplan, and A. E. Nelson, *Ann. Rev. Nucl. Part. Sci.* **43**, 27 (1993), and references therein.
[20] G. F. Giudice, *Phys. Rev. D* **45**, 3177 (1992); J. R. Espinosa, M. Quirós, and F. Zwirner, *Phys. Lett. B* **307**, 106 (1993).
[21] A. Abramovici *et al.*, *Science* **256**, 325 (1992). For detailed consideration of LIGO's capacity to detect stochastic backgrounds, see R. F. Michelson, *Mon. Not. R. Astr. Soc.* **227**, 993 (1987); N. L. Christensen, *Phys. Rev. D* **46**, 5250 (1993); and E. Flanagan, *Phys. Rev. D* **48**, 2389 (1993).
[22] J. E. Faller *et al.*, *Adv. Space Res.* **9**, 107 (1989).
[23] R. Courant and K. O. Friedrichs, *Supersonic Flows and Shock Waves* (Interscience Publishers, Inc., London, 1948).
[24] A. Linde, *Nucl. Phys.* **B216**, 421 (1983).
[25] M. Dine, R. Leigh, P. Huet, A. Linde and D. Linde, *Phys. Rev. D* **46**, 550 (1992).
[26] D.A. Kirzhnits, *JETP Lett.* **15**, 529 (1972); D. A. Kirzhnits and A. D. Linde, *Phys. Lett. B* **72**, 471 (1972); G. Anderson and L. Hall, *Phys. Rev. D* **45**, 2685 (1992).

FIGURES

FIG. 1. Velocity $\xi_d$ of propagation of detonation front as a function of $\alpha$.

FIG. 2. Fluid velocity for a detonation as a function of $\xi = r/t$ for: (a) $\alpha = 0.01$ (solid curve); (b) $\alpha = 1.0$ (dot-dash curve); and (c) $\alpha = 100$ (dashed curve).

FIG. 3. Enthalpy density, $w(\xi)$, divided by the enthalpy density $w_1$ outside the bubble, for a detonation as a function of $\xi$ for: (a) $\alpha = 0.01$ (solid curve); (b) $\alpha = 1.0$ (dot-dash curve); and (c) $\alpha = 100$ (dashed curve).

FIG. 4. Stress-energy density, $T(\xi) = wv^2\gamma^2$, for a detonation as a function of $\xi$ for: (a) $\alpha = 0.01$ (solid curve); (b) $\alpha = 1.0$ (dot-dash curve); and (c) $\alpha = 100$ (dashed curve).

FIG. 5. The fraction $\kappa$ of vacuum energy that goes into kinetic energy of the fluid in a detonation as a function of $\alpha$. The solid line is a numerical calculation; the dashed line is the analytic fit given by Eq. (22).

FIG. 6. The fraction of total energy (within an arbitrary volume) that is radiated into gravity waves by colliding bubbles as a function of bubble expansion velocity.

FIG. 7. The energy per octave radiated in gravity waves for a phase transition with spherical bubbles expanding at velocity $v$, for $v = 0.2$, $v = 0.4$, $v = 0.6$, $v = 0.8$, and $v = 1.0$.

FIG. 8. Fluid velocity for a deflagration as a function of $\xi = r/t$ for $v_2 = 0.1$ and $v_0 = 0.01$.

FIG. 9. Fluid velocity for a deflagration as a function of $\xi = r/t$ for $v_2 = 0.5$ and $v_0 = 0.45$ (solid curve) and $v_2 = 0.1$ and $v_0 = 0.09$ (dashed curve).

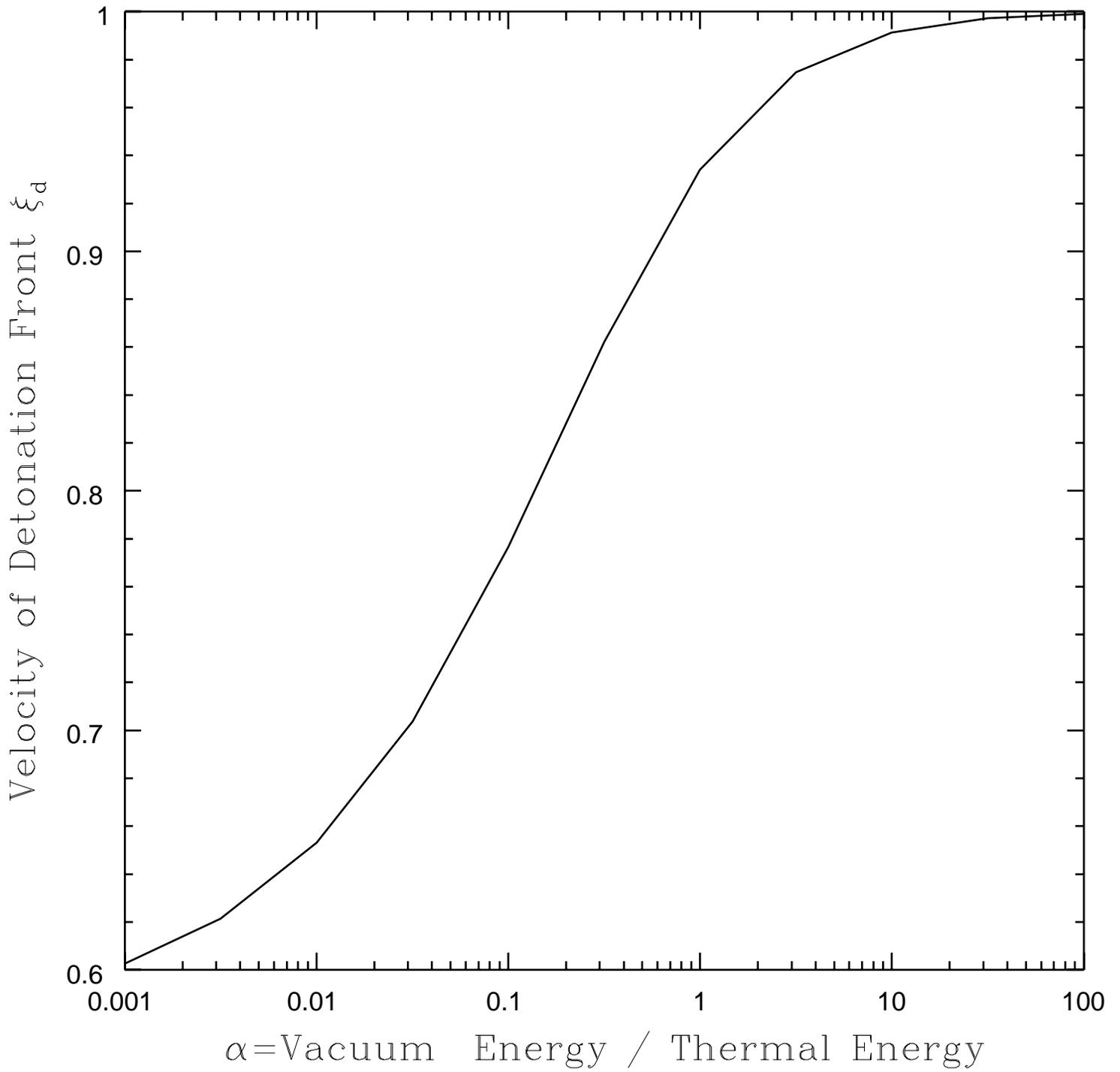

Figure 1

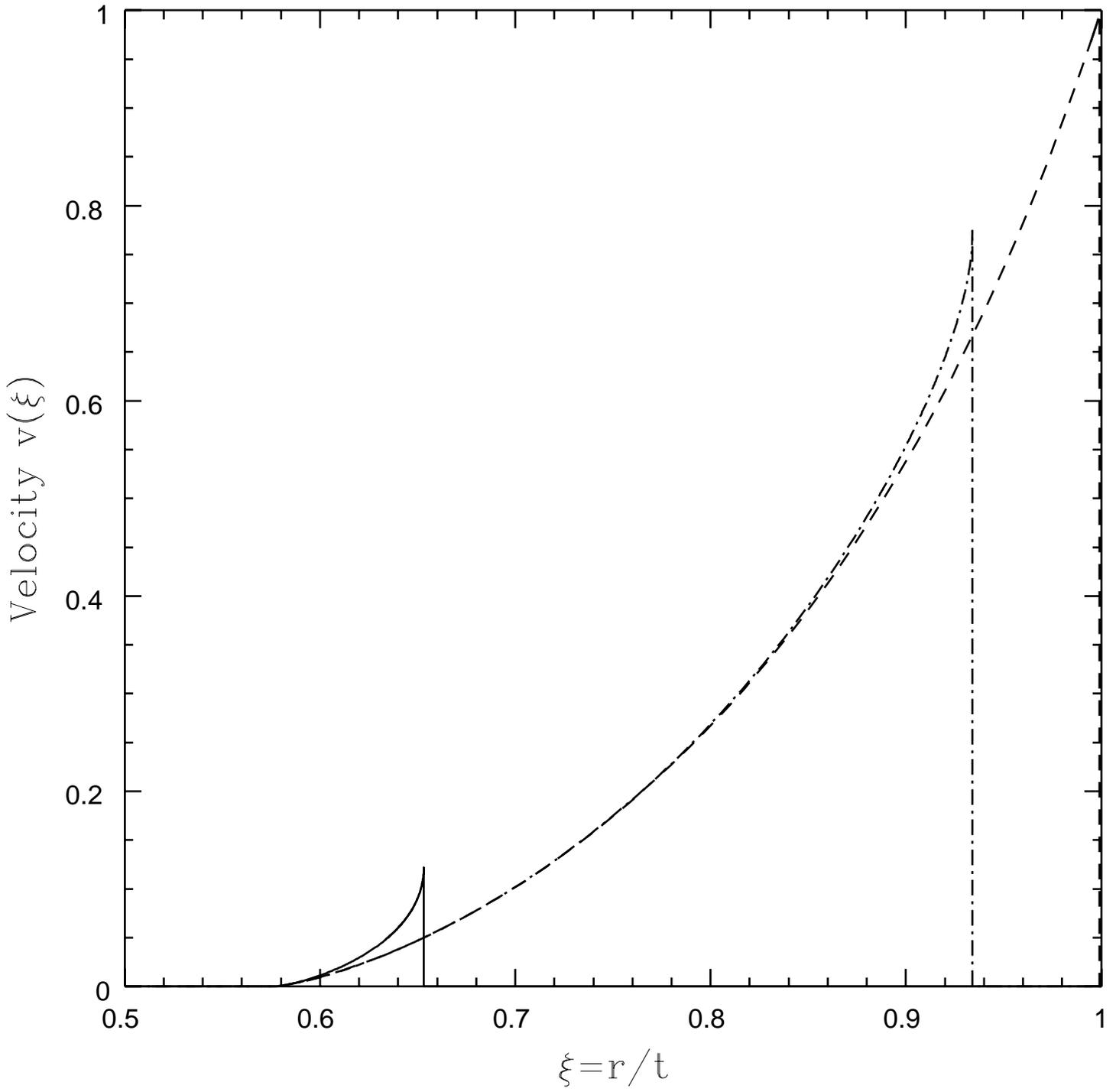

Figure 2

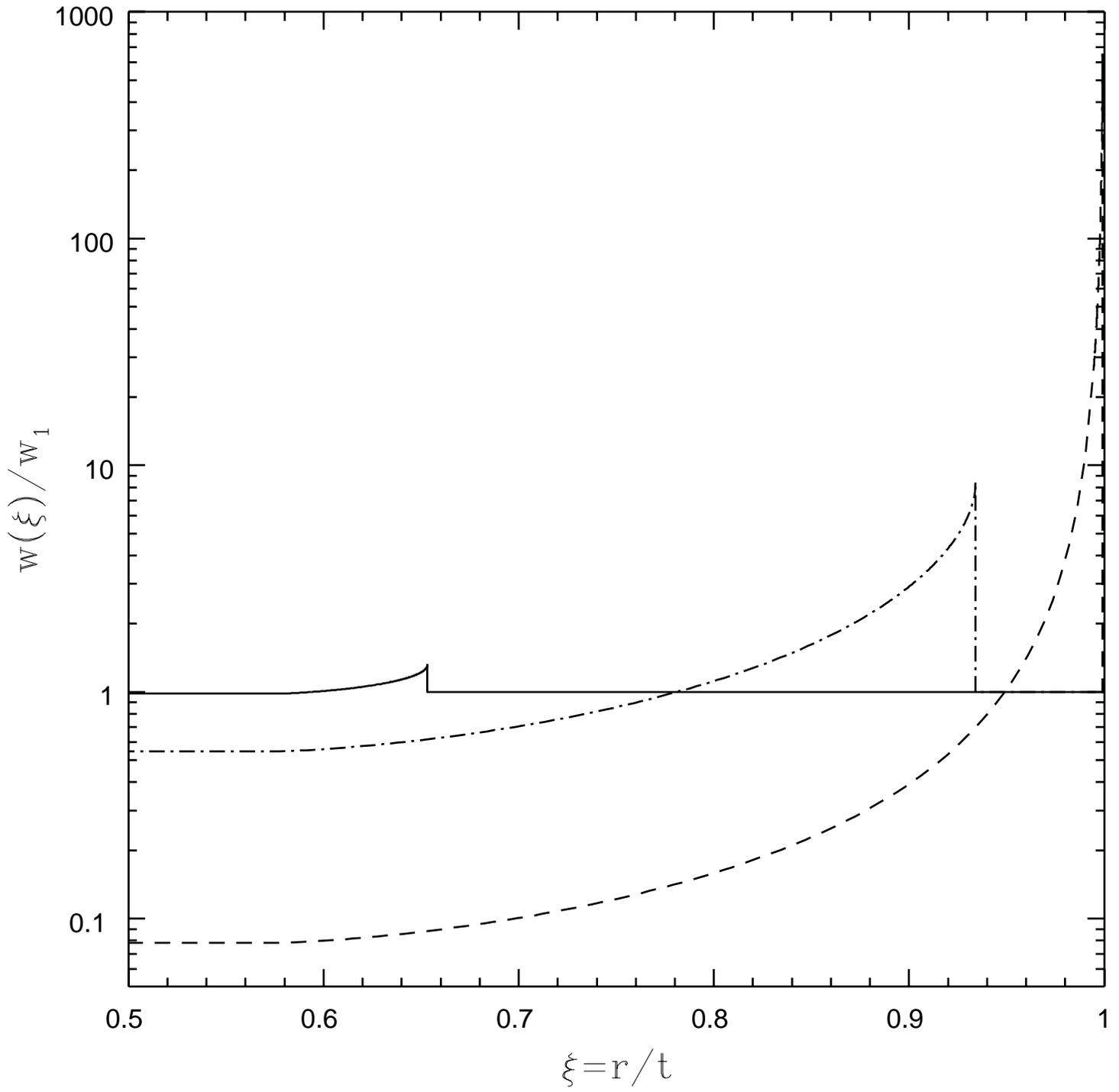

Figure 3

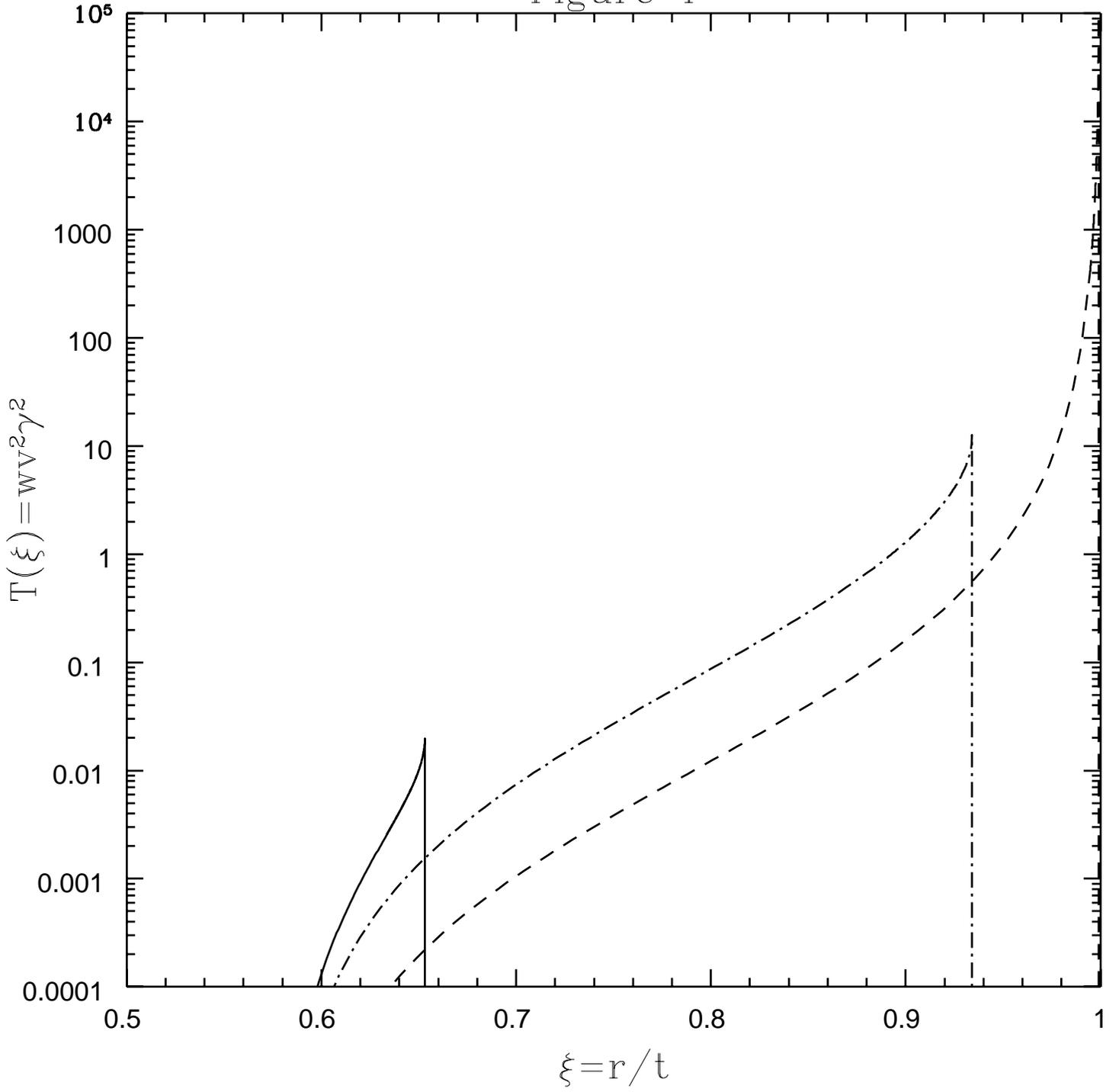

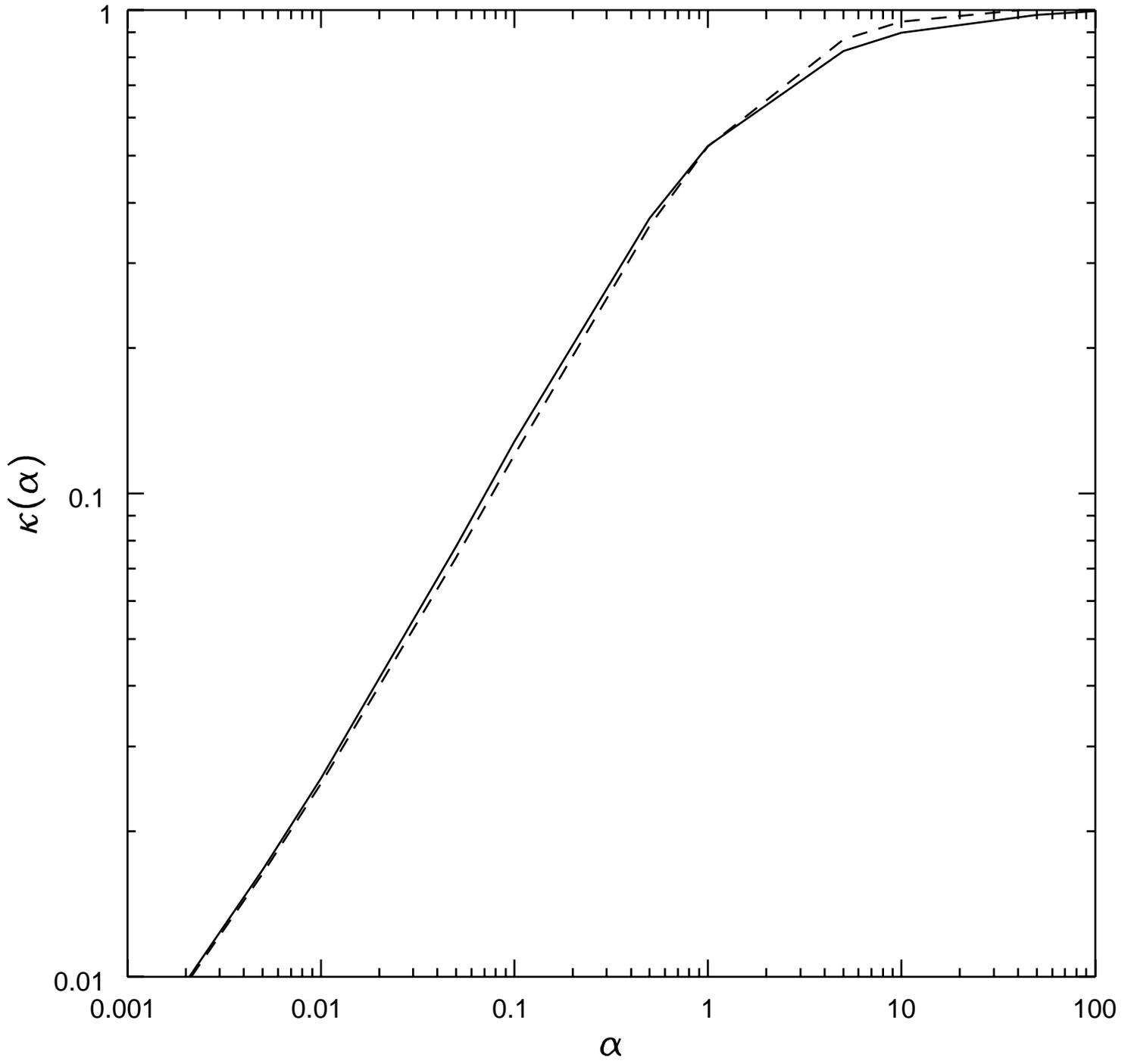

Figure 5

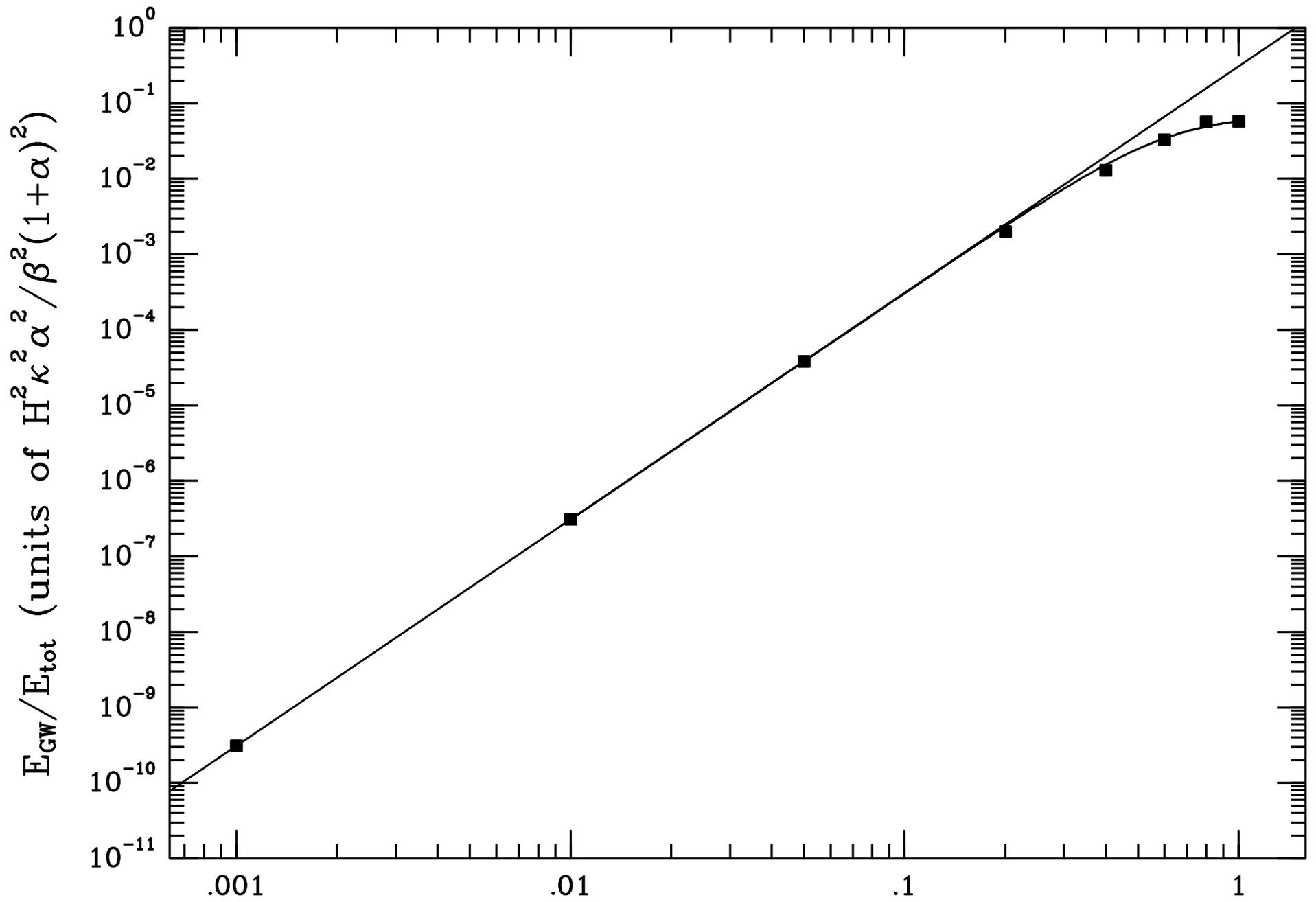

Figure 6

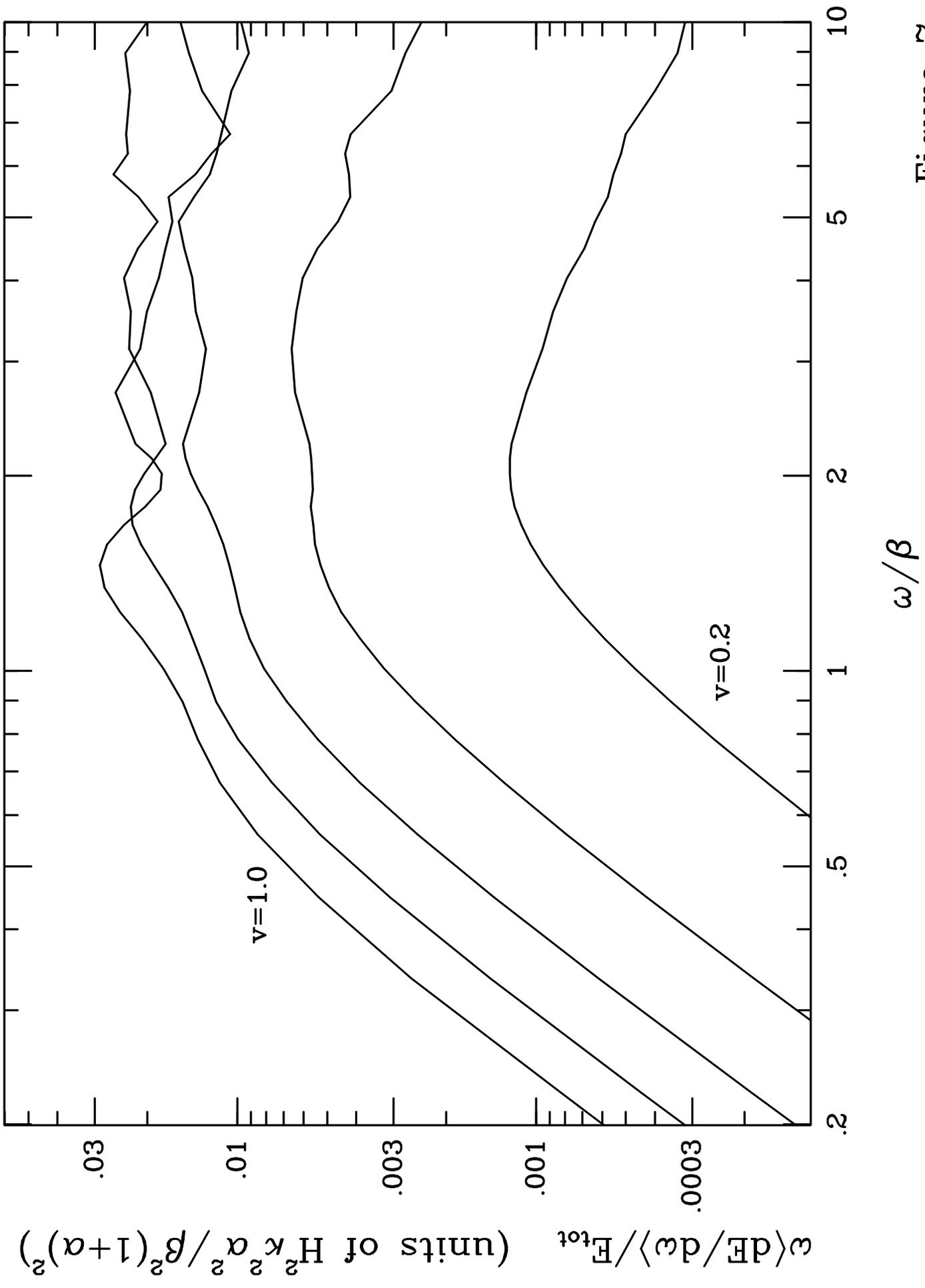

Figure 7

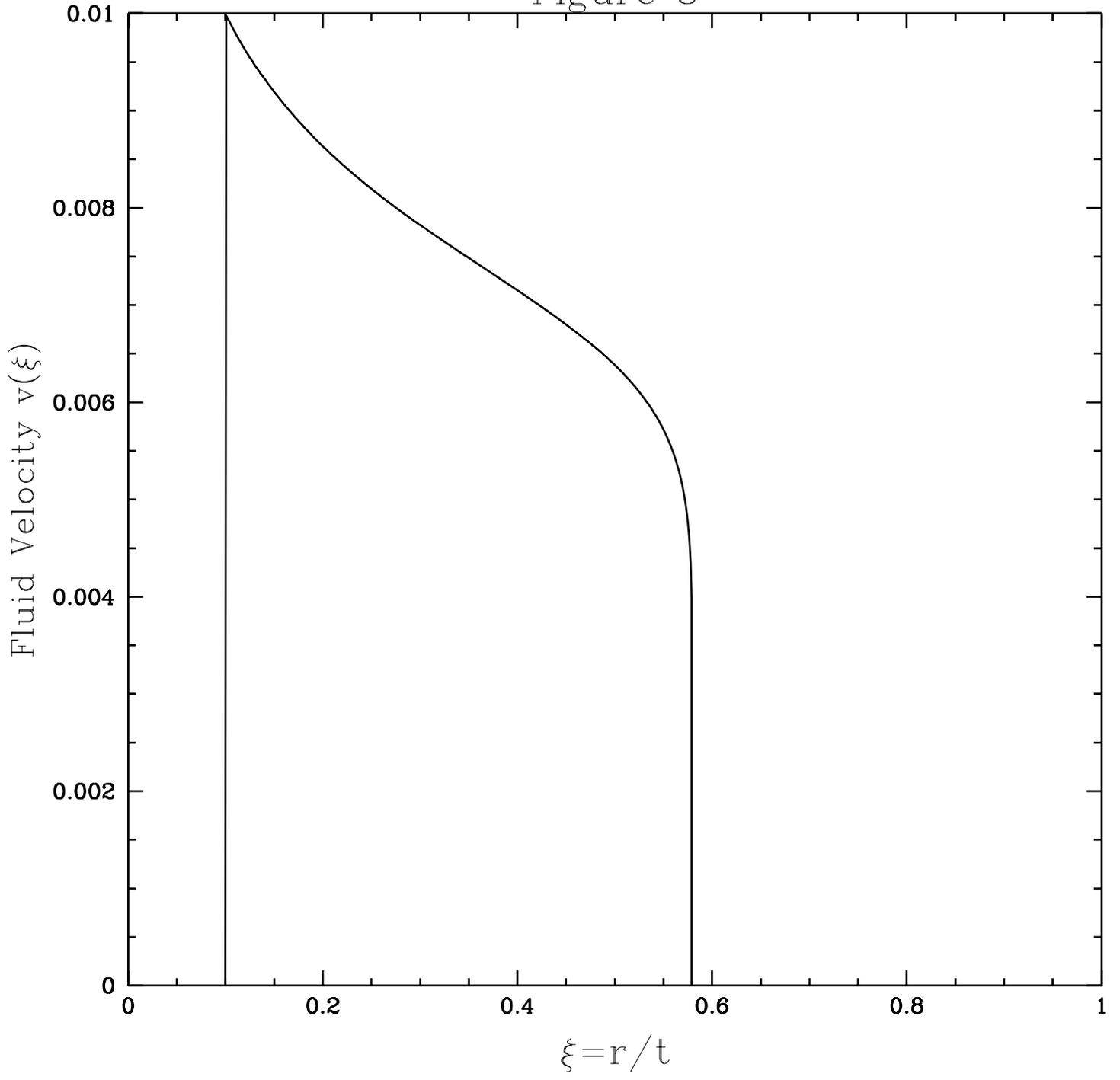

Figure 8

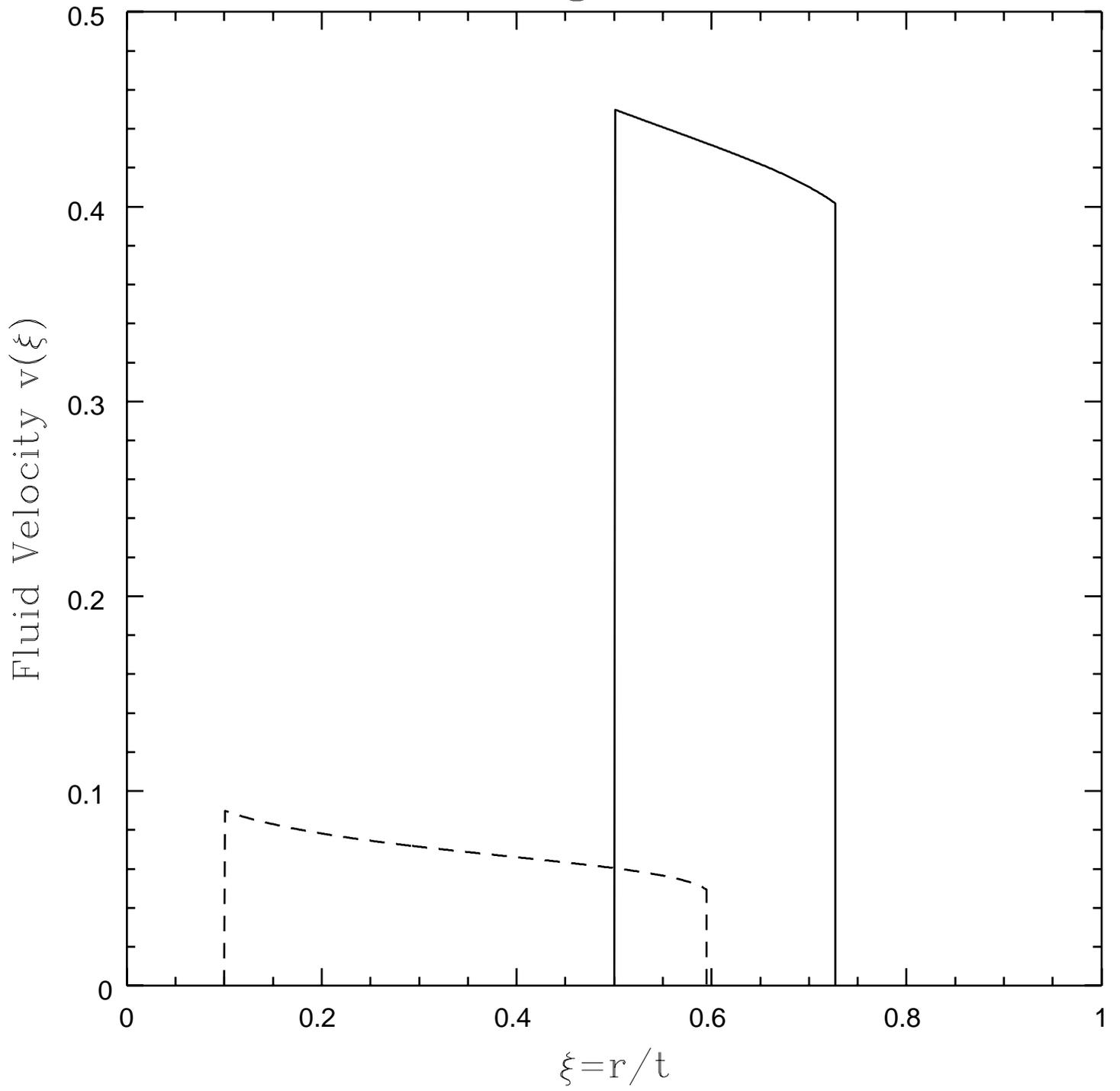

Figure 9